\documentclass{cargese}\usepackage{graphicx}

\let\footnote\savefootnote
\let\footnotetext\savefootnotetext 
\let\footnoterule\savefootnoterule 
\normallatexbib

\begin{document}

\articletitle[Cosmological Perturbations]
{Theory of Cosmological Perturbations and \\ 
Applications to Superstring Cosmology}


\author{Robert H. Brandenberger}



\affil{Physics Department, McGill University\\
3600 rue Universit\'e\\
Montreal, QC, H3A 2T8, CANADA\\
and\\
Brown University Physics Department\\
Providence, RI 02912, USA}    

\email{rhb@hep.physics.mcgill.ca}




\begin{abstract}

The theory of cosmological perturbations is the main tool
which connects theories of the early Universe (based on new
fundamental physics such as string theory) with cosmological
observations. In these lectures, I will provide an introduction
to this theory, beginning with an overview of the Newtonian
theory of fluctuations, moving on to the analysis of fluctuations
in the realm of classical general relativity, and culminating
with a discussion of the quantum theory of cosmological
perturbations. I will illustrate the formalism with
applications to inflationary cosmology. I will review
the basics of inflationary cosmology and discuss why -
through the evolution of fluctuations - inflation may provide
a way of observationally testing Planck-scale physics.

\end{abstract}

\section{Introduction}

Recent years have provided a wealth of observational data
about the cosmos. We have high resolution maps of the
anisotropies in the temperature of the cosmic microwave
background (CMB) \cite{WMAP}, surveys of the large-scale structure
(LSS) - the distribution of
galaxies in three-dimensional space - are increasing in
size and in accuracy (see e.g. \cite{2dF} and \cite{SDSS}),
and new techniques which will allow us to measure the
distribution of the dark matter are being pioneered.
All of this data involves small deviations of the cosmos
from homogeneity and isotropy. The cosmological observations
reveal that the Universe has non-random fluctuations on
all scales smaller than the present Hubble radius.

Parallel to this spectacular progress in observational cosmology,
new cosmological scenarios have emerged within which it is
possible to explain the origin of non-random inhomogeneities
by means of causal physics. The scenario which has attracted
most attention is inflationary cosmology \cite{Guth,Lindebook},
according to which there was a period in the early Universe in
which space was expanding at an accelerated rate. However, there
are also alternative proposals \cite{PBB,EKP} in which our
current stage of cosmological expansion is preceded by a phase
of contraction. These scenarios have in common the fact that for
scales of cosmological interest today, although their physical
wavelength is larger than the Hubble length during most of
the history of the universe, it is smaller than
the Hubble radius at very early times, thus in principle allowing
for a causal origin of the cosmological fluctuations.

In order to connect theories of fundamental physics providing an
origin of perturbations with the data on the late time universe,
one must be able to evolve cosmological fluctuations from earliest
times to today. Since on large scales (scales larger than about
$10$ Mpc - $1$ Mpc being roughly three million light years) 
the relative density fluctuations are smaller than one
today, and since these relative fluctuations grow in time as
a consequence of gravitational instability, they were smaller
than one throughout their history - at least in a universe which
is always expanding. Thus, it is reasonable to expect that a
linearized analysis of the fluctuations will give reliable results.
 
In most current models of the very early universe it is assumed that the
perturbations originate as quantum vacuum fluctuations. Thus,
quantum mechanics is important. On the other hand, since for most of
the history of the universe the wavelengths corresponding to scales
of cosmological interest today were larger than the Hubble radius,
it is crucial to consider the general relativistic theory
of fluctuations. Hence, a quantum theory of general relativistic
fluctuations is required. At the level of the linearized theory of
cosmological perturbations, a unified quantum theory of the formation
and evolution of fluctuations exists - and this will be the main
topic of these lectures. Note that all of the conceptual problems
of merging quantum mechanics and general relativity have been
thrown out by hand by restricting attention to the linearized
analysis. The question of ``linearization stability'' of the
system, namely whether the solutions of the linearized equations
of motion in fact correspond to the linearization of solutions of
the full equations, is a deep one and will not be addressed here
(see e.g. \cite{linstab} for discussions of this issue).

Inflationary cosmology is at the present time the most successful
framework of connecting physics of the very early universe with
the present structure (although alternatives such as the Pre-Big-Bang
\cite{PBB} and Ekpyrotic \cite{EKP} scenarios have been proposed and
may turn out to be successful as well).
I will thus begin these lectures with a review of inflationary cosmology,
and of scalar field-driven models for inflation. Next, I will
provide a detailed discussion of the theory of cosmological
perturbations, beginning with the Newtonian theory, moving on to
the classical general relativistic analysis, and ending with the
quantum theory of cosmological perturbations \footnote{These sections
are an updated version of the lecture notes \cite{Cancun}.} In
the final sections of these notes, I will return to inflationary
cosmology, and focus on some important conceptual problems which are 
not addressed in
current realizations of inflation in which the accelerated expansion
of space is driven by a scalar matter field. Addressing these
conceptual problems is a challenge and great opportunity for string
theory. Since cosmological inflation leads to a quasi-exponential increase
in the wavelength of inhomogeneities, it provides a microscope
with which string-scale physics can in principle be probed in 
current observations. I will conclude these lectures with a discussion
of this ``window of opportunity'' for string theory which inflation
provides \cite{RHBrev}. 

\section{Overview of Inflationary Cosmology}

To establish our notation and framework, we
will be taking the background space-time to be homogeneous
and isotropic, with a metric given by
\begin{equation} \label{background}
ds^2 \, = \, dt^2 - a(t)^2 d{\bf x}^2 \, ,
\end{equation}
where $t$ is physical time, $d{\bf x}^2$ is the Euclidean metric
of the spatial hypersurfaces (here taken for simplicity to be
spatially flat), and $a(t)$ is the scale factor. The scale
factor determines the {\it Hubble expansion rate} via 
\begin{equation}
H(t) \, = \, {{\dot{a}} \over a}(t) \, .
\end{equation}
The coordinates ${\bf x}$ used above are {\it comoving} coordinates,
coordinates painted onto the expanding spatial hypersurfaces.

In standard big bang cosmology, the universe is decelerating, i.e.
${\ddot a} < 0$. As a consequence, the {\it Hubble radius}
\begin{equation}
l_H(t) \, = \, H^{-1}(t)
\end{equation}
is increasing in comoving coordinates. As will be explained later
mathematically, the Hubble radius is the maximal distance that 
microphysics can act coherently over a Hubble expansion time -
in particular it is the maximal distance on which any causal process could
create fluctuations. If the universe were decelerating forever,
then scales of cosmological interest today would have had a 
wavelength larger than the Hubble radius at all early times. This
gives rise to the {\it fluctuation problem} for Standard Big Bang (SBB)
cosmology, namely the problem that there cannot be any causal process which
at early time creates perturbations on scales which are being
probed in current LSS and CMB observations \footnote{The reader is
encouraged to find the hole in this argument, and is referred to
\cite{ShellVil,HK,RHBtoprev} for the answer.}. 

The idea of inflationary cosmology is to assume that there was
a period in the very early Universe during which the scale factor was
accelerating, i.e. ${\ddot a} > 0$. This implies that the Hubble
radius was shrinking in comoving coordinates, or, equivalently, that fixed
comoving scales were ``exiting'' the Hubble radius. In the simplest models of
inflation, the scale factor increases nearly exponentially.
\begin{figure}
\centering
\includegraphics[height=6cm]{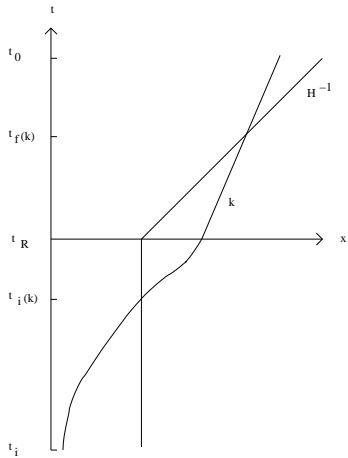}
\caption{Space-time diagram (sketch) showing the evolution
of scales in inflationary cosmology. The vertical axis is
time, and the period of inflation lasts between $t_i$ and
$t_R$, and is followed by the radiation-dominated phase
of standard big bang cosmology. During exponential inflation,
the Hubble radius $H^{-1}$ is constant in physical spatial coordinates
(the horizontal axis), whereas it increases linearly in time
after $t_R$. The physical length corresponding to a fixed
comoving length scale labelled by its wavenumber $k$ increases
exponentially during inflation but increases less fast than
the Hubble radius (namely as $t^{1/2}$), after inflation.}
\label{fig:1}       
\end{figure}
As illustrated in Figure (\ref{fig:1}), 
the basic geometry of inflationary cosmology
provides a solution of the fluctuation problem. As long as the phase
of inflation is sufficiently long, all length scales within our present
Hubble radius today originate at the beginning of inflation with a 
wavelength smaller than the Hubble radius at that time. Thus, it
is possible to create perturbations locally using physics obeying
the laws of special relativity (in particular causality). As will be
discussed later, it is quantum vacuum fluctuations of matter fields
and their associated curvature perturbations which are responsible for
the structure we observe today.

Postulating a phase of inflation in the very early universe also solves
the {\it horizon problem} of the SBB, namely it explains why the causal horizon
at the time $t_{rec}$ when photons last scatter is larger than the radius of the
past light cone at $t_{rec}$, the part of the last scattering surface which is 
visible today in CMB experiments. Inflation also explains the near flatness of
the universe: in a decelerating universe spatial flatness is an unstable fixed
point of the dynamics, whereas in an accelerating universe it becomes an
attractor. Another important aspect of the inflationary solution of the 
{\it flatness problem} is that inflation exponentially increases the volume 
of space. Without inflation, it is not possible that a Planck scale universe 
at the Planck time evolves into a sufficiently large universe today.

Let us now consider how it is possible to obtain a phase of cosmological
inflation. We will assume that space-time is to be described using the
equations of General Relativity \footnote{Note, however, that the first model
of exponential expansion of space \cite{Starob} made use of a higher derivative
gravitational action.}. In this case, the dynamics of the scale factor
$a(t)$ is determined by the Friedmann-Robertson-Walker (FRW) equations
\begin{equation} \label{FRW1}
({{\dot a} \over a})^2 \, = \, 8 \pi G \rho 
\end{equation}
and
\begin{equation} \label{FRW2}
{{\ddot a} \over a} \, = \, - 4 \pi G (\rho + 3p)
\end{equation}
where for simplicity we have omitted the contributions of spatial curvature
(since spatial curvature is diluted during inflation) and of the cosmological
constant (since any small cosmological constant which might be present today
has no effect in the early Universe since the associated energy density does
not increase when going into the past). In the above, $\rho$ and $p$ denote
the energy density and pressure, respectively. From (\ref{FRW2}) it is clear
that in order to obtain an accelerating universe, matter with sufficiently
negative pressure
\begin{equation}
p \, < \, - {1 \over 3} \rho
\end{equation}
is required. Exponential inflation is obtained for $p = - \rho$.

Conventional perfect fluids have positive semi-definite pressure and thus
cannot yield inflation. In addition, we know that a description of matter in
terms of classical perfect fluids must break down at early times. An improved
description of matter will be given in terms of quantum fields. Scalar
matter fields are special in that they allow at the level of a renormalizable
action the presence of a potential energy term. The energy density and
pressure of a scalar field $\varphi$ with canonically normalized action
\footnote{See \cite{kinflation} for a discussion of fields with non-canonical
kinetic terms.}
\begin{equation} \label{scalarlag}
{\cal L} \, = \, \sqrt{-g}
\bigl[{1 \over 2} \partial_{\mu} \varphi \partial^{\mu} \varphi
- V(\varphi)\bigr] \,  
\end{equation}
(where Greek indices are space-time indices, $V(\varphi)$ is the
potential energy density, and $g$ is the determinant of the metric) 
are given by
\begin{eqnarray}
\rho \, &=& \, {1 \over 2} ({\dot \varphi})^2 + 
{1 \over 2} a^{-2} (\nabla \varphi)^2 + V(\varphi) \nonumber \\
p \, &=& \, {1 \over 2} ({\dot \varphi})^2 - 
{1 \over 6} a^{-2} (\nabla \varphi)^2
- V(\varphi) \, .
\end{eqnarray}
Thus, it is possible to obtain an almost exponentially expanding universe 
provided the scalar field configuration \footnote{The scalar field
yielding inflation is called the {\it inflaton}.} satisfies
\begin{eqnarray}
{1 \over 2} (\nabla_p \varphi)^2 \, &\ll& \, V(\varphi) \, , \label{gradcond} \\
{1 \over 2} ({\dot \varphi})^2 \, &\ll& \, V(\varphi) \, . \label{tempcond}
\end{eqnarray}
In the above, $\nabla_p \equiv a^{-1} \nabla$ is the gradient with respect
to physical as opposed to comoving coordinates.
Since spatial gradients redshift as the universe expands, the first condition
will (for single scalar field models) always be satisfied if it is satisfied at
the initial time \footnote{In fact, careful studies \cite{Kung}
show that since the gradients decrease even in a non-inflationary 
backgrounds, they can become subdominant even if they are not initially 
subdominant.}. It is the second condition which is harder to satisfy. In
particular, this condition is in general not preserved in time even it is
initially satisfied. 

It is sufficient to obtain a period of cosmological inflation that
the {\it slow-roll conditions} for $\varphi$ are satisfied. Recall that
the equation of motion for a homogeneous scalar field in a cosmological
space-time is (as follows from (\ref{scalarlag})) is
\begin{equation} \label{scalareom}
{\ddot \varphi} + 3 H {\dot \varphi} \, = \, - V^{\prime}(\varphi) \, ,
\end{equation}
where a prime indicates the derivative with respect to $\varphi$. In order
that the scalar field roll slowly, it is necessary that
\begin{equation} \label{roll}
{\ddot \varphi} \, \ll \, 3 H {\dot \varphi}
\end{equation}
such that the first term in the scalar field equation of motion 
(\ref{scalareom}) is negligible. In this case, the condition (\ref{tempcond})
becomes
\begin{equation} \label{SRcond1}
({{V^{\prime}} \over V})^2 \, \ll \, 48 \pi G \,
\end{equation}
and (\ref{roll}) becomes
\begin{equation} \label{SRcond2}
{{V^{\prime \prime}} \over V} \, \ll \, 24 \pi G \, .
\end{equation}

In the initial model of inflation using scalar fields (``old inflation'' 
\cite{Guth}), it was assumed that $\varphi$ was initially in a false
vacuum with large potential energy. Hence, the conditions for inflation
are trivially satisfied. To end inflation, a quantum tunneling event from
the false vacuum to the true vacuum \cite{Coleman} was invoked (see e.g.
\cite{RMP} for a pedagogical review). This model, however, has a graceful 
exit problem since the tunneling leads to an initially microscopical bubble
of the true vacuum which cannot grow to encompass our presently observed 
universe - the flatness problem of SBB cosmology in a new form. Hence,
attention shifted to models in which the scalar field $\varphi$ is slowly
rolling during inflation.

There are many models of scalar field-driven inflation. Many of them can be
divided into three groups \cite{Kolb}: small-field inflation, large-field 
inflation and hybrid inflation. {\it Small-field inflationary models} are based
on ideas from spontaneous symmetry breaking in particle physics. We take the
scalar field to have a potential of the form
\begin{equation} \label{Mexican}
V(\varphi) \, = \, {1 \over 4} \lambda (\varphi^2 - \sigma^2)^2 \, ,
\end{equation}
where $\sigma$ can be interpreted as a symmetry breaking scale, and
$\lambda$ is a dimensionless coupling constant. The hope of initial
small-field models (``new inflation'' \cite{new}) was that the scalar
field would begin rolling close to its symmetric point $\varphi = 0$,
where thermal equilibrium initial conditions would localize it in the
early universe. At sufficiently high temperatures, $\varphi = 0$ 
is a stable ground state of the one-loop finite temperature effective 
potential $V_T(\varphi)$ (see e.g. \cite{RMP} for a review). 
Once the temperature drops to a value 
smaller than the critical temperature $T_c$, $\varphi = 0$ turns into 
an unstable local maximum of $V_T(\varphi)$, and $\varphi$ is free to 
roll towards a ground state of the zero temperature potential (\ref{Mexican}). 
The direction of the initial rolling is triggered by quantum fluctuations. 
The reader can
easily check that for the potential (\ref{Mexican}) the slow-roll conditions
cannot be satisfied if $\sigma \ll m_{pl}$, where $m_{pl}$ is the
Planck mass which is related to $G$. 
If the potential is modified to a Coleman-Weinberg \cite{CW} form 
\begin{equation} \label{CWpot}
V(\varphi) \, = \, {{\lambda} \over 4} \varphi^4
+ {{\lambda^2} \over {44 \pi^2}} \varphi^4 
\bigl[ {\rm ln} ({{\varphi^2} \over {M^2}}) - {{25} \over 6} \bigr]
\end{equation}
(where $M$ denotes some renormalization scale) 
then the slow-roll conditions can be satisfied. However, this
corresponds to a severe fine-tuning of the shape of the potential.
A further problem for most small-field models of inflation (see 
e.g. \cite{Goldwirth} for a review) is that the slow-roll trajectory
is not an attractor in phase space. In order to end up close to the
slow-roll trajectory, the initial field velocity must be constrained
to be very small. This {\it initial condition problem} of small-field
models of inflation effects a number of recently proposed brane inflation
scenarios, see e.g. \cite{GhazalScott} for a discussion. 

There is another reason for abandoning small-field inflation models: in
order to obtain a sufficiently small amplitude of density fluctuations,
the interaction coefficients of $\varphi$ must be very small (this
problem is discussed at length towards the end of these lectures). In
particular, this makes it inconsistent to assume that $\varphi$ started
out in thermal equilibrium. In the absence of thermal equilibrium, the
phase space of initial conditions is much larger for large values of
$\varphi$. 

This brings us to the discussion of large-field inflation
models, initially proposed in \cite{chaotic} under the name
``chaotic inflation''. The simplest example is
provided by a massive scalar field with potential
\begin{equation}
V(\varphi) \, = \, {1 \over 2} m^2 \varphi^2 \, ,
\end{equation}
where $m$ is the mass. It is assumed that the scalar field rolls towards
the origin from large values of $|\varphi|$. It is a simple exercise 
for the reader to verify that the slow-roll conditions (\ref{SRcond1}) and
(\ref{SRcond2}) are satisfied provided
\begin{equation}
|\varphi| \, > \, {1 \over {\sqrt{12 \pi}}} m_{pl} \, .
\end{equation}
Values of $|\varphi|$ comparable or greater than $m_{pl}$ are also
required in other realizations of large-field inflation. Hence, one
may worry whether such a toy model can consistently be embedded in
a realistic particle physics model, e.g. supergravity. In many such
models $V(\varphi)$ receives supergravity-induced correction terms
which destroy the flatness of the potential for $|\varphi| > m_{pl}$.
However, as recently discussed in \cite{Linderev}, if the flatness
of the potential is protected by some symmetry, then it can survive
inclusion of the correction terms. As will be discussed later, a value
of $m \sim 10^{13}$GeV is required in order to obtain the observed
amplitude of density fluctuations. Hence, the configuration space of
field values with $|\varphi| > m_{pl}$ but $V(\varphi) < m_{pl}^4$ is
huge. It can also be verified that the slow-roll trajectory is a local
attractor in field initial condition space \cite{Kung}, even including
metric fluctuations at the perturbative level \cite{Feldman}.

With two scalar fields it is possible to construct a class of models
which combine some of the nice features of large-field inflation
(large phase space of initial conditions yielding inflation) and of
small-field inflation (better contact with conventional particle physics).
These are models of hybrid inflation \cite{hybrid}. To give a prototypical
example, consider two scalar fields $\varphi$ and $\chi$ with a potential
\begin{equation} \label{hybridpot}
V(\varphi, \chi) \, = \, {1 \over 4} \lambda_{\chi} (\chi^2 - \sigma^2)^2
+ {1 \over 2} m^2 \varphi^2 - {1 \over 2} g^2 \varphi^2 \chi^2 \, .
\end{equation}
In the absence of thermal equilibrium, it is natural to assume that 
$|\varphi|$ begins at large values, values for which the effective mass
of $\chi$ is positive and hence $\chi$ begins at $\chi = 0$. The parameters
in the potential (\ref{hybridpot}) are now chosen such that $\varphi$
is slowly rolling for values of $|\varphi|$ somewhat smaller than $m_{pl}$,
but that the potential energy for these field values is dominated by
the first term on the right-hand side of (\ref{hybridpot}). The reader
can easily verify that for this model it is no longer required to have
values of $|\varphi|$ greater than $m_{pl}$ in order to obtain slow-rolling
\footnote{Note that the slow-roll conditions (\ref{SRcond1}) and (\ref{SRcond2})
were derived assuming that $H$ is given by the contribution of
$\varphi$ to $V$ which is not the case here.}
The field $\varphi$ is slowly rolling whereas the potential energy is
determined by the contribution from $\chi$. Once $|\varphi|$ drops to the
value
\begin{equation}
|\varphi_c| \, = \, {{{\sqrt{\lambda_{\chi}}}} \over g} \sigma 
\end{equation}
the configuration $\chi = 0$ becomes unstable and decays to its ground
state $|\chi| = \sigma$, yielding a graceful exit from inflation.
Since in this example the ground state of $\chi$ is not unique, there
is the possibility of the formation of topological defects at the end
of inflation (see \cite{ShellVil,HK,RHBtoprev} for reviews of topological
defects in cosmology, and the lectures by Polchinski at this school
\cite{PolCar} for a discussion of how this
scenario arises in brane inflation models).

After the slow-roll conditions break down, the period of inflation ends,
and the inflaton begins to oscillate around its ground state.
Due to couplings of $\varphi$ to other matter fields, the energy of the
universe, which at the end of the period of inflation is stored completely
in $\varphi$, gets transferred to the matter fields of the particle
physics Standard Model. Initially, the energy transfer was described
perturbatively \cite{DolLin,AFW}. Later, it was realized \cite{TB,KLS,STB,KLS2}
that through a parametric resonance instability, particles are very
rapidly produced, leading to a fast energy transfer (``preheating'').
The quanta later thermalize, and thereafter the universe evolves as described
by SBB cosmology.

After this review of inflationary cosmology (see e.g. \cite{Lythbook} for
a more complete recent review), we turn to the discussion of the main
success of inflationary cosmology, namely the fact that it provides
a causal mechanism for generating small inhomogeneities. The reader is
referred to \cite{MFB} for a comprehensive analysis of this theory of
cosmological perturbations.

\section{Newtonian Theory of Cosmological Perturbations}
\label{rhbsec:1}

\subsection{Introduction}

The growth of density fluctuations is a consequence of the
purely attractive nature of the gravitational force. Imagine (first
in a non-expanding background)
a density excess $\delta \rho$ localized about some point ${\bf x}$ in space.
This fluctuation produces an attractive force which pulls
the surrounding matter towards ${\bf x}$. The magnitude of this
force is proportional to $\delta \rho$. Hence, by Newton's
second law
\begin{equation} \label{rhbeq1}
\ddot{\delta \rho} \, \sim \, G \delta \rho \, ,
\end{equation}
where $G$ is Newton's gravitational constant. Hence, there is
an exponential instability of flat space-time to the development
of fluctuations. 

Obviously, in General Relativity it is inconsistent to consider
density fluctuations in a non-expanding background. If we
consider density fluctuations in an expanding background,
then the expansion of space leads to a friction term in (\ref{rhbeq1}).
Hence, instead of an exponential instability to the development of
fluctuations, the growth rate of fluctuations in an expanding Universe
will be as a power of time. It is crucial to determine what this power
is and how it depends both on the background cosmological expansion rate
and on the length scale of the fluctuations.

Note that in the following two subsections ${\bf x}$ will denote
the physical coordinates, and ${\bf q}$ the comoving ones.
The materials covered in this section are discussed in several
excellent textbooks on cosmology, e.g. in 
\cite{Weinberg,Peebles,Padmanabhan,Peacock}.
 
\subsection{Perturbations about Minkowski Space-Time}

To develop some physical intuition, we first consider the
evolution of hydrodynamical matter fluctuations in a fixed
non-expanding background.  

In this context, matter is described
by a perfect fluid, and gravity by the Newtonian gravitational
potential $\varphi$. The fluid variables are the energy density
$\rho$, the pressure $p$, the fluid velocity ${\bf v}$, and
the entropy density $S$. The basic hydrodynamical equations
are
\begin{eqnarray} \label{rhbeq2}
\dot{\rho} + \nabla \cdot (\rho {\bf v}) & = & 0 \nonumber \\
\dot{{\bf v}} + ({\bf v} \cdot \nabla) {\bf v} + {1 \over {\rho}} \nabla p
+ \nabla \varphi & = & 0 \nonumber \\
\nabla^2 \varphi & = & 4 \pi G \rho \\
\dot{S} + ({\bf v} \cdot \nabla) S & = & 0 \nonumber \\
p & = & p(\rho, S) \, . \nonumber
\end{eqnarray}
The first equation is the continuity equation, the second is the Euler
(force) equation, the third is the Poisson equation of Newtonian gravity,
the fourth expresses entropy conservation, and the last describes
the equation of state of matter. The derivative with respect to time
is denoted by an over-dot.

The background is given by the background energy density $\rho_o$, 
the background
pressure $p_0$, vanishing velocity, constant gravitational potential
$\varphi_0$ and constant entropy density $S_0$. As mentioned above, it
does not satisfy the background Poisson equation.

The equations for cosmological perturbations are obtained by perturbing
the fluid variables about the background,
\begin{eqnarray} \label{rhbeq3}
\rho & = & \rho_0 + \delta \rho \nonumber \\
{\bf v} & = & \delta {\bf v} \nonumber \\
p & = & p_0 + \delta p \\
\varphi & = & \varphi_0 + \delta \varphi \nonumber \\
S & = & S_0 + \delta S \, , \nonumber
\end{eqnarray}
where the fluctuating fields $\delta \rho, \delta {\bf v}, \delta p,
\delta \varphi$ and $\delta S$ are functions of space and time, by
inserting these expressions into the basic hydrodynamical equations
(\ref{rhbeq2}), by linearizing, and by combining the resulting equations
which are of first order in time. We get the following
differential equations for the energy density fluctuation $\delta \rho$
and the entropy perturbation $\delta S$
\begin{eqnarray} \label{rhbeq4}
\ddot{\delta \rho} - c_s^2 \nabla^2 \delta \rho - 4 \pi G \rho_0 \delta \rho
& = & \sigma \nabla^2 \delta S \\
\dot \delta S \ & = & 0 \, , \nonumber
\end{eqnarray}
where the variables $c_s^2$ and $\sigma$ describe the equation of state
\begin{equation} \label{pressurepert}
\delta p \, = \, c_s^2 \delta \rho + \sigma \delta S
\end{equation}
with 
\begin{equation}
c_s^2 \, = \, \bigl({{\delta p} \over {\delta \rho}}\bigr)_{|_{S}}
\end{equation}
denoting the square of the speed of sound.

Since the equations are linear, we can work in Fourier space. Each
Fourier component $\delta \rho_k(t)$ of the fluctuation field 
$\delta \rho({\bf x}, t)$ 
\begin{equation}
\delta \rho ({\bf x}, t) \, = \, 
\int e^{i {\bf k} \cdot {\bf x}} \delta \rho_k(t)
\end{equation}
evolves independently.

The fluctuations can be classified as follows: If 
$\delta S$ vanishes, we have {\bf adiabatic} fluctuations. If
the $\delta S$ is non-vanishing but 
$\dot{\delta \rho} = 0$, we speak of an {\bf entropy} fluctuation.

The first conclusions we can draw from the basic perturbation
equations (\ref{rhbeq4}) are that \\
1) entropy fluctuations do not grow, \\
2) adiabatic fluctuations are time-dependent, and \\
3) entropy fluctuations seed an adiabatic mode.

Taking a closer look at the equation of motion for
$\delta \rho$, we see that the third term on the left hand side
represents the force due to gravity, a purely attractive force 
yielding an instability of flat space-time to the development of
density fluctuations (as discussed earlier, see (\ref{rhbeq1})).
The second term on the left hand side of (\ref{rhbeq4}) represents
a force due to the fluid pressure which tends to set up pressure waves.
In the absence of entropy fluctuations, the evolution of $\delta \rho$
is governed by the combined action of both pressure and gravitational
forces.

Restricting our attention to adiabatic fluctuations, we see from
(\ref{rhbeq4}) that there is a critical wavelength, the Jeans length,
whose wavenumber $k_J$ is given by
\begin{equation} \label{Jeans}
k_J \, = \, \bigl({{4 \pi G \rho_0} \over {c_s^2}}\bigr)^{1/2} \, .
\end{equation}
Fluctuations with wavelength longer than the Jeans length ($k \ll k_J$)
grow exponentially
\begin{equation} \label{expgrowth}
\delta \rho_k(t) \, \sim \, e^{\omega_k t} \,\, {\rm with} \,\,
\omega_k \sim 4 (\pi G \rho_0)^{1/2}
\end{equation}
whereas short wavelength modes ($k \gg k_J$) oscillate with 
frequency $\omega_k \sim c_s k$. Note that the value of the
Jeans length depends on the equation of state of the background.
For a background dominated by relativistic radiation, the Jeans
length is large (of the order of the Hubble radius $H^{-1}(t)$), 
whereas for pressure-less matter it goes to zero.

\subsection{Perturbations about an Expanding Background}

Let us now improve on the previous analysis and study Newtonian
cosmological fluctuations about an expanding background. In this
case, the background equations are consistent (the non-vanishing
average energy density leads to cosmological expansion). However,
we are still neglecting general relativistic effects (the 
fluctuations of the metric). Such effects
turn out to be dominant on length scales larger than the Hubble
radius $H^{-1}(t)$, and thus the analysis of this section is
applicable only to smaller scales. 

The background cosmological model is given by the energy density
$\rho_0(t)$, the pressure $p_0(t)$, and the recessional velocity
${\bf v}_0 = H(t) {\bf x}$, where ${\bf x}$ is the Euclidean spatial
coordinate vector (``physical coordinates''). The space- and time-dependent
fluctuating fields are defined as in the previous section:
\begin{eqnarray} \label{fluctansatz2}
\rho(t, {\bf x}) & = & \rho_0(t) \bigl(1 + \delta_{\epsilon}(t, {\bf x}) 
\bigr)\nonumber \\
{\bf v}(t, {\bf x}) & = & {\bf v}_0(t, {\bf x}) + \delta {\bf v}(t, {\bf x}) 
\\
p(t, {\bf x}) & = & p_0(t) + \delta p(t, {\bf x}) \, , \nonumber 
\end{eqnarray}
where $\delta_{\epsilon}$ is the fractional energy density perturbation
(we are interested in the fractional rather than in the absolute energy
density fluctuation!), and the pressure perturbation $\delta p$ is
defined as in (\ref{pressurepert}). In addition, there is the
possibility of a non-vanishing entropy perturbation defined as in
(\ref{rhbeq3}).

We now insert this ansatz into the basic hydrodynamical equations 
(\ref{rhbeq2}), linearize in the perturbation variables, and combine
the first order differential equations 
for $\delta_{\epsilon}$ and $\delta p$ into a single second order
differential equation for $\delta \rho_{\epsilon}$. The result simplifies
if we work in ``comoving coordinates'' ${\bf q}$ which are the coordinates
painted onto the expanding background, i.e. 
\begin{equation}
{\bf x}(t) \, = \, a(t) {\bf q}(t) \, .
\end{equation}
After some algebra, we obtain the following equation
which describes the time evolution of density fluctuations:
\begin{equation} \label{Newtoneq}
\ddot{\delta_{\epsilon}} + 2 H \dot{\delta_{\epsilon}} 
- {{c_s^2} \over {a^2}} \nabla_q^2 \delta_{\epsilon} 
- 4 \pi G \rho_0 \delta_{\epsilon} \, 
= \, {{\sigma} \over {\rho_0 a^2}} \delta S \, ,
\end{equation}
where the subscript $q$ on the $\nabla$ operator indicates that derivatives
with respect to comoving coordinates are used.
In addition, we have the equation of entropy conservation
\begin{equation}
\dot{\delta S} \, = \, 0 \, .
\end{equation}

Comparing with the equations (\ref{rhbeq4}) obtained in the absence of
an expanding background, we see that the only difference is the presence
of a Hubble damping term in the equation for $\delta_{\epsilon}$. This
term will moderate the exponential instability of the background to
long wavelength density fluctuations. In addition, it will lead to a
damping of the oscillating solutions on short wavelengths. More specifically,
for physical wavenumbers $k_p \ll k_J$ (where $k_J$ is again given by
(\ref{Jeans})), and in a matter-dominated background cosmology, the
general solution of (\ref{Newtoneq}) in the absence of any entropy
fluctuations is given by
\begin{equation} \label{Newtonsol}
\delta_k(t) \, = \, c_1 t^{2/3} + c_2 t^{-1} \, ,
\end{equation}
where $c_1$ and $c_2$ are two constants determined by the initial
conditions, and we have dropped the subscript $\epsilon$ in expressions
involving $\delta_{\epsilon}$. 
There are two fundamental solutions, the first is a
growing mode with $\delta_k(t) \sim a(t)$, the second a decaying
mode with $\delta_k(t) \sim t^{-1}$.
On short wavelength, one obtains damped oscillatory motion:
\begin{equation} \label{Newtonsolosc}
\delta_k(t) \, \sim \, a^{-1/2}(t) exp \bigl( \pm i c_s k \int dt' a^{-1}(t')
\bigr) \, .
\end{equation}

As a simple application of the Newtonian equations for cosmological
perturbations derived above, let us compare the predicted cosmic
microwave background (CMB) anisotropies in a spatially
flat Universe with only baryonic matter - Model A -
to the corresponding anisotropies
in a flat Universe with mostly cold dark matter (pressure-less non-baryonic
dark matter) - Model B. We start with the observationally known amplitude
of the relative density fluctuations today (time $t_0$), 
and we use the fact that
the amplitude of the CMB anisotropies on the angular scale $\theta(k)$
corresponding to the comoving wavenumber $k$ is set by the primordial
value of the gravitational potential $\phi$ - introduced in the
following section - which in turn is related to the primordial value of the
density fluctuations at Hubble radius crossing (and {\bf not}
to its value of the time $t_{rec}$). See e.g. Chapter 17 of \cite{MFB}).

In Model A, the dominant component of the pressure-less matter is
coupled to radiation between $t_{eq}$ and $t_{rec}$, the time of
last scattering. Thus, the Jeans length is comparable to the Hubble
radius. Therefore, for comoving galactic scales, $k \gg k_J$ in this
time interval, and thus the fractional density contrast decreases
as $a(t)^{-1/2}$. In contrast, in Model B, the dominant component of
pressure-less matter couples only weakly to radiation, and hence
the Jeans length is negligibly small. Thus, in Model B, the
relative density contrast grows as $a(t)$ between $t_{eq}$ and $t_{rec}$.
In the time interval $t_{rec} < t < t_0$, the fluctuations scale
identically in Models A and B. Summarizing,
we conclude, working backwards in time from a fixed amplitude
of $\delta_k$ today, that the amplitudes of $\delta_k(t_{eq})$ in Models
A and B (and thus their primordial values) are related by
\begin{equation}
\delta_k(t_{eq})|_{A} \, \simeq \, 
\bigl({{a(t_{rec})} \over {a(t_{eq})}} \bigr) \delta_k(t_{eq})|_{B} \, .
\end{equation}
Hence, in Model A (without non-baryonic dark matter) the CMB anisotropies
are predicted to be a factor of about 30 larger 
\cite{SW} than in Model B, way
in excess of the recent observational results. This is one of the
strongest arguments for the existence of non-baryonic dark matter. Note that
the precise value of the enhancement factor depends on the value of the
cosmological constant $\Lambda$ - 
the above value holds for $\Lambda = 0$. 

\subsection{Characterizing Perturbations}

Let us consider perturbations on a fixed
comoving length scale given by a comoving wavenumber $k$. 
The corresponding physical length increases
as $a(t)$. This is to be compared to the Hubble radius $H^{-1}(t)$
which scales as $t$ provided $a(t)$ grows as a power of $t$. In
the late time Universe, $a(t) \sim t^{1/2}$ in the radiation-dominated
phase (i.e. for $t < t_{eq}$, and $a(t) \sim t^{2/3}$ in the 
matter-dominated period ($t_{eq} < t < t_0$). 
Thus, we see that at sufficiently early times, all comoving scales
had a physical length larger than the Hubble radius. If we consider 
large cosmological scales (e.g. those corresponding to the observed
CMB anisotropies or to galaxy clusters), the time $t_H(k)$ of 
``Hubble radius crossing'' (when the physical length was equal to the
Hubble radius) was in fact later than $t_{eq}$. As we will see
in later sections, the time of Hubble radius crossing plays an
important role in the evolution of cosmological perturbations.

Cosmological fluctuations can be described either in position space
or in momentum space. In position space, we compute the root mean
square mass fluctuation $\delta M / M(k, t)$ in a sphere of radius
$l = 2 \pi / k$ at time $t$. A scale-invariant spectrum of fluctuations is
defined by the relation
\begin{equation} \label{scaleinv}
{{\delta M} \over M}(k, t_H(k)) \, = \, {\rm const.} \, .
\end{equation}
Such a spectrum was first suggested by Harrison \cite{Harrison}
and Zeldovich \cite{Zeldovich} as a reasonable choice for the
spectrum of cosmological fluctuations. We can introduce the ``spectral
index'' $n$ of cosmological fluctuations by the relation
\begin{equation} \label{specindex}
\bigl({{\delta M} \over M}\bigr)^2(k, t_H(k)) \, \sim \, k^{n - 1} \, ,
\end{equation}
and thus a scale-invariant spectrum corresponds to $n = 1$.

To make the transition to the (more frequently used) momentum space
representation, we Fourier decompose the fractional spatial density
contrast
\begin{equation} \label{Fourier}
\delta_{\epsilon}({\bf x}, t) \, = \, 
\int d^3k {\tilde{\delta_{\epsilon}}}({\bf k}, t) e^{i {\bf k} \cdot {\bf x}}
\, .
\end{equation}
The {\bf power spectrum} $P_{\delta}$ of density fluctuations is defined by
\begin{equation} \label{densspec}
P_{\delta}(k) \, = \, k^3 |{\tilde{\delta_{\epsilon}}}(k)|^2 \, ,
\end{equation}
where $k$ is the magnitude of ${\bf k}$, and we have assumed for simplicity
a Gaussian distribution of fluctuations in which the amplitude of the
fluctuations only depends on $k$.

We can also define the power spectrum of the gravitational potential $\varphi$:
\begin{equation} \label{gravspec}
P_{\varphi}(k) \, = \, k^3 |{\tilde{\delta \varphi}}(k)|^2 \, .
\end{equation}
These two power spectra are related by the Poisson equation (\ref{rhbeq2})
\begin{equation} \label{relspec}
P_{\varphi}(k) \, \sim \, k^{-4} P_{\delta}(k) \, .
\end{equation}

In general, the condition of scale-invariance is expressed in momentum
space in terms of the power spectrum evaluated at a fixed time. To obtain
this condition, we first use the time dependence of the 
fractional density fluctuation from (\ref{Newtonsol}) to determine
the mass fluctuations at a fixed time $t > t_H(k) > t_{eq}$ (the last
inequality is a condition on the scales considered)
\begin{equation} \label{timerel}
\bigl({{\delta M} \over M}\bigr)^2(k, t) \, = \,
\bigl({t \over {t_H(k)}}\bigr)^{4/3} 
\bigl({{\delta M} \over M}\bigr)^2(k, t_H(k)) \, .
\end{equation}
The time of Hubble radius crossing is given by
\begin{equation} \label{Hubble}
a(t_H(k)) k^{-1} \, = \, 2 t_H(k) \, ,
\end{equation} 
and thus
\begin{equation} \label{Hubble2}
t_H(k)^{1/2} \, \sim \, k^{-1} \, .
\end{equation}
Inserting this result into (\ref{timerel}) making use of (\ref{specindex})
we find 
\begin{equation} \label{spec2}
\bigl({{\delta M} \over M}\bigr)^2(k, t) \, \sim \, k^{n + 3} \, .
\end{equation}
Since, for reasonable values of the index of the power spectrum, 
$\delta M / M (k, t)$ is dominated by the Fourier modes with
wavenumber $k$, we find that (\ref{spec2}) implies
\begin{equation} \label{spec3}
|{\tilde{\delta_{\epsilon}}}|^2 \, \sim \, k^{n} \, ,
\end{equation}
or, equivalently,
\begin{equation} \label{spec4}
P_\varphi(k) \, \sim \, k^{n - 1} \, .
\end{equation}

\section{Relativistic Theory of Cosmological Fluctuations}

\subsection{Introduction}

The Newtonian theory of cosmological fluctuations discussed in the
previous section breaks down on scales larger than the Hubble radius
because it neglects perturbations of the metric, and because on large
scales the metric fluctuations dominate the dynamics.

Let us begin with a heuristic argument to show why metric fluctuations
are important on scales larger than the Hubble radius. For such
inhomogeneities, one should be able to approximately describe the
evolution of the space-time by applying the first 
FRW equation (\ref{FRW1}) of homogeneous
and isotropic cosmology to the local Universe (this approximation is
made more rigorous in \cite{Afshordi}).
Based on this equation, a large-scale fluctuation of the
energy density will lead to a fluctuation (``$\delta a$'') of
the scale factor $a$ which grows in time. This is due to the fact
that self gravity amplifies fluctuations even on length scales $\lambda$
greater than the Hubble radius.

This argument is made rigorous in the following analysis of cosmological
fluctuations in the context of general relativity, where both metric
and matter inhomogeneities are taken into account. We will consider
fluctuations about a homogeneous and isotropic background cosmology,
given by the metric (\ref{background}), which can be written in
conformal time $\eta$ (defined by $dt = a(t) d\eta$) as
\begin{equation} \label{background2}
ds^2 \, = \, a(\eta)^2 \bigl( d\eta^2 - d{\bf x}^2 \bigr) \, .
\end{equation}

The theory of cosmological perturbations is based on expanding the Einstein
equations to linear order about the background metric. The theory was
initially developed in pioneering works by Lifshitz \cite{Lifshitz}. 
Significant progress in the understanding of the physics of cosmological
fluctuations was achieved by Bardeen \cite{Bardeen} who realized the
importance of subtracting gauge artifacts (see below) from the
analysis (see also \cite{PV}). The following discussion is based on
Part I of the comprehensive review article \cite{MFB}. Other reviews - in
some cases emphasizing different approaches - are 
\cite{Kodama,Ellis,Hwang,Durrer}.

\subsection{Classifying Fluctuations}

The first step in the analysis of metric fluctuations is to
classify them according to their transformation properties
under spatial rotations. There are scalar, vector and second rank
tensor fluctuations. In linear theory, there is no coupling
between the different fluctuation modes, and hence they evolve
independently (for some subtleties in this classification, see
\cite{Stewart}). 

We begin by expanding the metric about the FRW background metric
$g_{\mu \nu}^{(0)}$ given by (\ref{background2}):
\begin{equation} \label{pertansatz}
g_{\mu \nu} \, = \, g_{\mu \nu}^{(0)} + \delta g_{\mu \nu} \, .
\end{equation}
The background metric depends only on time, whereas the metric
fluctuations $\delta g_{\mu \nu}$ depend on both space and time.
Since the metric is a symmetric tensor, there are at first sight
10 fluctuating degrees of freedom in $\delta g_{\mu \nu}$.

There are four degrees of freedom which correspond to scalar metric
fluctuations (the only four ways of constructing a metric from
scalar functions):
\begin{equation} \label{scalarfl}
\delta g_{\mu \nu} \, = \, a^2 \left(
\begin{array} {cc}
2 \phi & -B_{,i} \\
-B_{,i} & 2\bigl(\psi \delta_{ij} - E_{,ij} \bigr) 
\end{array}
\right) \, ,
\end{equation}
where the four fluctuating degrees of freedom are denoted (following
the notation of \cite{MFB}) $\phi, B, E$, and $\psi$, a comma denotes
the ordinary partial derivative (if we had included spatial curvature
of the background metric, it would have been the covariant derivative
with respect to the spatial metric), and $\delta_{ij}$ is the Kronecker
symbol.

There are also four vector degrees of freedom of metric fluctuations,
consisting of the four ways of constructing metric fluctuations from
three vectors:
\begin{equation} \label{vectorfl}
\delta g_{\mu \nu} \, = \, a^2 \left(
\begin{array} {cc}
0 & -S_i \\
-S_i & F_{i,j} + F_{j,i} 
\end{array}
\right) \, ,
\end{equation}
where $S_i$ and $F_i$ are two divergence-less vectors (for a vector
with non-vanishing divergence, the divergence contributes to the
scalar gravitational fluctuation modes).

Finally, there are two tensor modes which correspond to the two
polarization states of gravitational waves:
\begin{equation} \label{tensorfl}
\delta g_{\mu \nu} \, = \, -a^2 \left(
\begin{array} {cc}
0 & 0 \\
0 & h_{ij} 
\end{array}
\right) \, ,
\end{equation}
where $h_{ij}$ is trace-free and divergence-less
\begin{equation}
h_i^i \, = \, h_{ij}^j \, = \, 0 \, .
\end{equation}

Gravitational waves do not couple at linear order to the matter
fluctuations. Vector fluctuations decay in an expanding background
cosmology and hence are not usually cosmologically important.
The most important fluctuations, at least in inflationary cosmology,
are the scalar metric fluctuations, the fluctuations which couple
to matter inhomogeneities and which are the relativistic generalization
of the Newtonian perturbations considered in the previous section.

\subsection{Gauge Transformation}

The theory of cosmological perturbations is at first sight complicated
by the issue of gauge invariance (at the final stage, however, we will
see that we can make use of the gauge freedom to substantially simplify
the theory). The coordinates $t, {\bf x}$ of space-time carry no
independent physical meaning. They are just labels to designate points
in the space-time manifold. By performing a small-amplitude 
transformation of the space-time coordinates
(called ``gauge transformation'' in the following), we can easily
introduce ``fictitious'' fluctuations in a homogeneous and isotropic
Universe. These modes are ``gauge artifacts''.

We will in the following take an ``active'' view of gauge transformation.
Let us consider two space-time manifolds, one of them a homogeneous
and isotropic Universe ${\cal M}_0$, the other a physical Universe 
${\cal M}$ with inhomogeneities. A choice of coordinates can be considered
to be a mapping ${\cal D}$ between the manifolds ${\cal M}_0$ and ${\cal M}$.
Let us consider a second mapping ${\tilde{\cal D}}$ which will map the
same point (e.g. the origin of a fixed coordinate system) in ${\cal M}_0$
into different points in ${\cal M}$. Using the inverse of these maps
${\cal D}$ and ${\tilde{\cal D}}$, we can assign two different sets of
coordinates to points in ${\cal M}$. 

Consider now a physical quantity $Q$ (e.g. the Ricci scalar)
on ${\cal M}$, and the corresponding
physical quantity $Q^{(0)}$ on ${\cal M}_0$ Then, in the first coordinate
system given by the mapping ${\cal D}$, the perturbation $\delta Q$ of
$Q$ at the point $p \in {\cal M}$ is defined by
\begin{equation}
\delta Q(p) \, = \, Q(p) - Q^{(0)}\bigl({\cal D}^{-1}(p) \bigr) \, .
\end{equation}
Analogously, in the second coordinate system given by ${\tilde{\cal D}}$,
the perturbation is defined by
\begin{equation}
{\tilde{\delta Q}}(p) \, = \, Q(p) - 
Q^{(0)}\bigl({\tilde{{\cal D}}}^{-1}(p) \bigr) \, .
\end{equation}
The difference
\begin{equation}
\Delta Q(p) \, = \, {\tilde{\delta Q}}(p) - \delta Q(p)
\end{equation}
is obviously a gauge artifact and carries no physical significance.

Some of the metric perturbation degrees of freedom introduced in
the first subsection are gauge artifacts. To isolate these, 
we must study how coordinate transformations act on the metric.
There are four independent gauge degrees of freedom corresponding
to the coordinate transformation
\begin{equation}
x^{\mu} \, \rightarrow \, {\tilde x}^{\mu} = x^{\mu} + \xi^{\mu} \, .
\end{equation}
The zero (time) component $\xi^0$ of $\xi^{\mu}$ leads to a scalar metric
fluctuation. The spatial three vector $\xi^i$ can be decomposed as
\begin{equation}
\xi^i \, = \, \xi^i_{tr} + \gamma^{ij} \xi_{,j}
\end{equation}
(where $\gamma^{ij}$ is the spatial background metric)
into a transverse piece $\xi^i_{tr}$ which has two degrees of freedom
which yield vector perturbations, and the second term (given by
the gradient of a scalar $\xi$) which gives
a scalar fluctuation. To summarize this paragraph, there are
two scalar gauge modes given by $\xi^0$ and $\xi$, and two vector
modes given by the transverse three vector $\xi^i_{tr}$. Thus,
there remain two physical scalar and two vector
fluctuation modes. The gravitational waves are gauge-invariant. 

Let us now focus on how the scalar gauge transformations (i.e. the
transformations given by $\xi^0$ and $\xi$) act on the scalar
metric fluctuation variables $\phi, B, E$, and $\psi$. An immediate
calculation yields:
\begin{eqnarray}
{\tilde \phi} \, &=& \, \phi - {{a'} \over a} \xi^0 - (\xi^0)^{'} \nonumber \\
{\tilde B} \, &=& \, B + \xi^0 - \xi^{'} \\
{\tilde E} \, &=& \, E - \xi \nonumber \\
{\tilde \psi} \, &=& \, \psi + {{a'} \over a} \xi^0 \, , \nonumber
\end{eqnarray}
where a prime indicates the derivative with respect to conformal time $\eta$.

There are two approaches to deal with the gauge ambiguities. The first is
to fix a gauge, i.e. to pick conditions on the coordinates which
completely eliminate the gauge freedom, the second is to work with a
basis of gauge-invariant variables.

If one wants to adopt the gauge-fixed approach, there are many
different gauge choices. Note that the often used synchronous gauge
determined by $\delta g^{0 \mu} = 0$ does not totally fix the
gauge. A convenient system which completely fixes the coordinates
is the so-called {\bf longitudinal} or {\bf conformal Newtonian gauge}
defined by $B = E = 0$.

If one prefers a gauge-invariant approach, there are many choices
of gauge-invariant variables. A convenient basis first introduced
by \cite{Bardeen} is the basis $\Phi, \Psi$ given by 
\begin{eqnarray} \label{givar}
\Phi \, &=& \, \phi + {1 \over a} \bigl[ (B - E')a \bigr]^{'} \\
\Psi \, &=& \, \psi - {{a'} \over a} (B - E') \, .
\end{eqnarray}
It is obvious from the above equations that the gauge-invariant
variables $\Phi$ and $\Psi$ coincide with the corresponding
diagonal metric perturbations $\phi$ and $\psi$ in longitudinal
gauge. 

Note that the variables defined above are gauge-invariant only
under linear space-time coordinate transformations. Beyond
linear order, the structure of perturbation theory becomes much
more involved. In fact, one can show \cite{SteWa} that the only
fluctuation variables which are invariant under all coordinate
transformations are perturbations of variables which are constant
in the background space-time.

\subsection{Equations of Motion}

We begin with the Einstein equations
\begin{equation} \label{Einstein}
G_{\mu\nu} \, = \, 8 \pi G T_{\mu\nu} \, , 
\end{equation}
where $G_{\mu\nu}$ is the Einstein tensor associated with the space-time
metric $g_{\mu\nu}$, and $T_{\mu\nu}$ is the energy-momentum tensor of matter,
insert the ansatz for metric and matter perturbed about a FRW 
background $\bigl(g^{(0)}_{\mu\nu}(\eta) ,\, \varphi^{(0)}(\eta)\bigr)$:
\begin{eqnarray} \label{pertansatz2}
g_{\mu\nu} ({\bf x}, \eta) & = & g^{(0)}_{\mu\nu} (\eta) + \delta g_{\mu\nu}
({\bf x}, \eta) \\
\varphi ({\bf x}, \eta) & = & \varphi_0 (\eta) + \delta \varphi
({\bf x}, \eta) \, , 
\end{eqnarray}
(where we have for simplicity replaced general matter by a scalar
matter field $\varphi$)
and expand to linear order in the fluctuating fields, obtaining the
following equations:
\begin{equation} \label{linein}
\delta G_{\mu\nu} \> = \> 8 \pi G \delta T_{\mu\nu} \, .
\end{equation}
In the above, $\delta g_{\mu\nu}$ is the perturbation in the metric and $\delta
\varphi$ is the fluctuation of the matter field $\varphi$.

Note that the components $\delta G^{\mu}_{\nu}$ and $\delta T^{\mu}_{\nu}$
are not gauge invariant. If we want to use the gauge-invariant approach,
we note \cite{MFB} that it is possible to construct a gauge-invariant tensor 
$\delta G^{(gi) \, \mu}_{\nu}$ via
\begin{eqnarray} \label{givar2}
\delta G^{(gi) \, 0}_{0} \, &\equiv& \, \delta G^{0}_{0}
+ (^{(0)}G^{' \, 0}_0)(B - E') \, \nonumber \\
\delta G^{(gi) \, 0}_{i} \, &\equiv& \, \delta G^{0}_{i}
+ (^{(0)}G^0_i - {1 \over 3} {}^{(0)}G^k_k)(B - E')_{,i} \, \\
\delta G^{(gi) \, i}_{j} \, &\equiv& \, \delta G^{i}_{j}
+ (^{(0)}G^{' \, i}_j)(B - E') \, , \nonumber
\end{eqnarray}
where ${}^(0)G^{\mu}_{\nu}$ denote the background values of the
Einstein tensor.
Analogously, a gauge-invariant linearized stress-energy tensor
$\delta T^{(gi) \, \mu}_{\nu}$
can be defined. In terms of these tensors, the gauge-invariant form of
the equations of motion for linear fluctuations reads
\begin{equation} \label{linefin}
\delta G_{\mu\nu}^{(gi)} \> = \> 8 \pi G \delta T_{\mu\nu}^{(gi)} \, .
\end{equation}
If we insert into this equation the ansatz for the general metric
and matter fluctuations (which depend on the gauge), 
only gauge-invariant combinations of the
fluctuation variables will appear.

In a gauge-fixed approach, one can start with the metric in
longitudinal gauge
\begin{equation} \label{longit}
ds^2 \, = \, a^2 \bigl[(1 + 2 \phi) d\eta^2
- (1 - 2 \psi)\gamma_{ij} dx^i dx^j \bigr] \,
\end{equation}
and insert this ansatz into the general perturbation equations
(\ref{linein}). The shortcut of inserting a restricted
ansatz for the metric into the action and deriving the full set
of variational equations is justified in this case. 

Both approaches yield the following set of equations of motion:
\begin{eqnarray} \label{perteom1}
- 3 {\cal H} \bigl( {\cal H} \Phi + \Psi^{'} \bigr) + \nabla^2 \Psi \,
&=& \, 4 \pi G a^2 \delta T^{(gi) \, 0}_0 \nonumber \\
\bigl( {\cal H} \Phi + \Psi^{'} \bigr)_{, i} \,
&=& 4 \pi G a^2 \delta T^{(gi) \, 0}_i \\
\bigl[ \bigl( 2 {\cal H}^{'} + {\cal H}^2 \bigr) \Phi + {\cal H} \Phi^{'}
+ \Psi^{''} + 2 {\cal H} \Psi^{'} \bigr] \delta^i_j && \nonumber \\
+ {1 \over 2} \nabla^2 D \delta^i_j - {1 \over 2} \gamma^{ik} D_{, kj} \,
&=& - 4 \pi G a^2 \delta T^{(gi) \, i}_j \, , \nonumber
\end{eqnarray}
where $D \equiv \Phi - \Psi$ and ${\cal H} = a'/a$. If we work in
longitudinal gauge, then $\delta T^{(gi) \, i}_j = \delta T^i_j$,
$\Phi = \phi$ and $\Psi = \psi$.

The first conclusion we can draw is that if no anisotropic stress
is present in the matter at linear order in fluctuating fields, i.e.
$\delta T^i_j = 0$ for $i \neq j$, then the two metric fluctuation
variables coincide:
\begin{equation} \label{constr}
\Phi \, = \, \Psi \, .
\end{equation}
This will be the case in most simple cosmological models, e.g. in
theories with matter described by a set of scalar fields with
canonical form of the action, and in the case of a perfect fluid
with no anisotropic stress.

Let us now restrict our attention to the case of matter described
in terms of a single scalar field $\varphi$ which can be 
expanded as
\begin{equation} \label{mfexp}
\varphi ({\bf x}, \eta) \, = \, \varphi_0(\eta) + \delta \varphi({\bf x}, \eta)
\end{equation}
in terms of background matter $\varphi_0$ and matter fluctuation 
$\delta \varphi({\bf x}, \eta)$. Then, in longitudinal gauge,
(\ref{perteom1}) reduce to the following
set of equations of motion (making use of (\ref{constr}))
\begin{eqnarray} \label{perteom2}
\nabla^2 \phi - 3 {\cal H} \phi^{'} - 
\bigl( {\cal H}^{'} + 2 {\cal H}^2 \bigr) \phi \, &=& \, 
4 \pi G \bigl( \varphi^{'}_0 \delta \varphi^{'} + 
V^{'} a^2 \delta \varphi \bigr) \nonumber \\
\phi^{'} + {\cal H} \phi \, &=& \, 4 \pi G \varphi^{'}_0 \delta \varphi \\
\phi^{''} + 3 {\cal H} \phi^{'} + 
\bigl( {\cal H}^{'} + 2 {\cal H}^2 \bigr) \phi \, &=& \, 
4 \pi G \bigl( \varphi^{'}_0 \delta \varphi^{'} - 
V^{'} a^2 \delta \varphi \bigr) \, , \nonumber
\end{eqnarray}
where $V^{'}$ denotes the derivative of $V$ with respect to $\varphi$.
These equations can be combined to give the following second order
differential equation for the relativistic potential $\phi$:
\begin{equation} \label{finaleom}
\phi^{''} + 2 \left( {\cal H} - 
{{\varphi^{''}_0} \over {\varphi^{'}_0}} \right) \phi^{'} - \nabla^2 \phi
+ 2 \left( {\cal H}^{'} - 
{\cal H}{{\varphi^{''}_0} \over {\varphi^{'}_0}} \right) \phi \, = \, 0 \, .
\end{equation}

This is the final result for the classical evolution of
cosmological fluctuations. First of all, we note the similarities with
the equation (\ref{Newtoneq}) obtained in the Newtonian theory.
The final term in (\ref{finaleom}) is the force due to gravity leading
to the instability, the second to last term is the pressure force
leading to oscillations (relativistic since we are considering matter
to be a relativistic field), and the second term is the Hubble friction
term. For each wavenumber there are two fundamental solutions. On
small scales ($k > H$), the solutions correspond to damped oscillations,
on large scales ($k < H$) the oscillations freeze out and the dynamics
is governed by the gravitational force competing with the Hubble friction
term. Note, in particular, how the Hubble radius naturally emerges as
the scale where the nature of the fluctuating modes changes from oscillatory
to frozen.

Considering the equation in a bit more detail, observe that if the
equation of state of the background is independent of time (which will be
the case if ${\cal H}^{'} = \varphi^{''}_0 = 0$), then in an
expanding background, the dominant mode of (\ref{finaleom}) is constant,
and the sub-dominant mode decays. If the equation of state is not constant,
then the dominant mode is not constant in time. Specifically, at the
end of inflation ${\cal H}^{'} < 0$, and this leads to a growth of 
$\phi$ (see the following subsection).

To study the quantitative implications of the equation of motion
(\ref{finaleom}), it is convenient to introduce \cite{BST,BK}
the variable $\zeta$ (which, up to correction term of the order
$\nabla^2 \phi$ which is unimportant for large-scale fluctuations,
is equal to the curvature perturbation ${\cal R}$ in comoving gauge
\cite{Lyth})
by
\begin{equation} \label{zetaeq}
\zeta \, \equiv \, \phi + {2 \over 3} 
{{\bigl(H^{-1} {\dot \phi} + \phi \bigr)} \over { 1 + w}} \, ,
\end{equation}
where
\begin{equation} \label{wvar}
w = {p \over {\rho}}
\end{equation}
characterizes the equation of state of matter. In terms of $\zeta$,
the equation of motion (\ref{finaleom}) takes on the form
\begin{equation}
{3 \over 2} {\dot \zeta} H (1 + w) \, = \, {\cal O}(\nabla^2 \phi) \, .
\end{equation}
On large scales, the right hand side of the equation is negligible,
which leads to the conclusion that large-scale cosmological fluctuations
satisfy
\begin{equation} \label{zetacons}
{\dot \zeta} (1 + w) \, = \, 0 .
\end{equation}
This implies that $\zeta$ is constant
except possibly if $1 + w = 0$ at some point in time during the cosmological 
evolution (which occurs during reheating in inflationary
cosmology if the inflaton field undergoes oscillations - see 
\cite{Fabio1} and \cite{BV,Fabio2} for discussions of the consequences
in single and double field inflationary models, respectively). 
In single matter field models it is indeed possible
to show that ${\dot \zeta} = 0$ on super-Hubble scales independent
of assumptions on the equation of state \cite{Weinberg2,Zhang}.
This ``conservation law'' makes it easy to relate
initial fluctuations to final fluctuations in inflationary cosmology,
as will be illustrated in the following subsection.

\subsection{Application to Inflationary Cosmology}
\label{Sec1}

Let us now return to the space-time sketch of the evolution of
fluctuations in inflationary cosmology - see Figure (\ref{fig:1}) - and use the
conservation law (\ref{zetacons}) - in the form
$\zeta = {\rm const}$ on large scales - to relate the amplitude
of $\phi$ at initial Hubble radius crossing during the inflationary
phase (at $t = t_i(k)$) with the amplitude at final Hubble radius
crossing at late times (at $t = t_f(k)$). Since both at early
times and at late times ${\dot \phi} = 0$ on super-Hubble scales
as the equation of state is not changing, (\ref{zetacons}) and (\ref{zetaeq})
lead to
\begin{equation} \label{inflcons}
\phi(t_f(k)) \, \simeq \, {{(1 + w)(t_f(k))} \over {(1 + w)(t_i(k))}}
\phi(t_i(k)) \, .
\end{equation} 

This equation will allow us to evaluate the amplitude of the
cosmological perturbations when they re-enter the Hubble radius
at time $t_f(k)$, under the assumption (discussed in detail in
the following section) that the origin of the primordial
fluctuations is quantum vacuum oscillations. 

The time-time perturbed Einstein equation (the first equation
of (\ref{perteom1})) relates the value of $\phi$ at initial
Hubble radius crossing to the amplitude of the fractional energy
density fluctuations. This, together with the fact that
the amplitude of the scalar matter field quantum vacuum fluctuations
is of the order $H$, yields
\begin{equation} \label{phiin}
\phi(t_i(k)) \, \sim \, H {{V^{'}} \over V}(t_i(k)) \, .
\end{equation}
In the late time radiation dominated phase, $w = 1/3$,
whereas during slow-roll inflation
\begin{equation} \label{win}
1 + w(t_i(k)) \, \simeq \, {{{\dot \varphi_0}^2} \over V}(t_i(k)) \, .
\end{equation}
Making, in addition, use of the slow roll conditions satisfied
during the inflationary period
\begin{eqnarray} \label{srcond}
H {\dot \varphi_0} \, &\simeq& \, - V^{'}  \nonumber \\
H^2 \, &\simeq& \, {{8 \pi G} \over 3} V \, ,
\end{eqnarray}
we arrive at the final result
\begin{equation} \label{final}
\phi(t_f(k)) \, \sim \, {{V^{3/2}} \over {V^{'} m_{pl}^3}}(t_i(k)) \, ,
\end{equation}
which gives the position space amplitude of cosmological
fluctuations on a scale labelled by the comoving wavenumber $k$
at the time when the scale re-enters the Hubble radius at
late times, a result first obtained in the case of the
Starobinsky model \cite{Starob} of inflation in \cite{Mukh1},
and later in the context of scalar field-driven inflation
in \cite{GuthPi,Starob4,Hawking,BST}.

In the case of slow roll inflation, the right hand side of
(\ref{final}) is, to a first approximation, independent of $k$,
and hence the resulting spectrum of fluctuations is nearly
scale-invariant.
 
\section{Quantum Theory of Cosmological Fluctuations}

\subsection{Overview}

As already mentioned in the previous
section, in many models of the very early Universe, in particular
in inflationary cosmology, but also in the Pre-Big-Bang and
in the Ekpyrotic scenarios, primordial inhomogeneities emerge
from quantum vacuum fluctuations on microscopic scales (wavelengths
smaller than the Hubble radius). The wavelength is then stretched
relative to the Hubble radius, becomes larger than the Hubble
radius at some time and then propagates on super-Hubble scales until
re-entering at late cosmological times. In the context of a Universe
with a de Sitter phase, the quantum origin of cosmological fluctuations
was first discussed in \cite{Mukh1} - see \cite{Lukash} for
a more general discussion of the quantum origin of fluctuations
in cosmology, and also \cite{Press,Sato} for
earlier ideas. In particular, Mukhanov \cite{Mukh1} and Press 
\cite{Press} realized that
in an exponentially expanding background, the curvature fluctuations
would be scale-invariant, and Mukhanov provided a quantitative
calculation which also yielded the logarithmic deviation from
exact scale-invariance. 

To understand the role
of the Hubble radius, consider the equation of a free scalar matter field
$\varphi$ on an unperturbed expanding background:
\begin{equation}
\ddot{\varphi} + 3 H \dot{\varphi} - {{\nabla^2} \over {a^2}} \varphi
\, = \, 0 \, .
\end{equation}
The second term on the left hand side of this equation leads to damping
of $\varphi$ with a characteristic decay rate given by $H$. As a
consequence, in the absence of the spatial gradient term, $\dot{\varphi}$
would be of the order of magnitude $H \varphi$. Thus, comparing the
second and the third term on the left hand side, we immediately
see that the microscopic (spatial gradient) term dominates on length
scales smaller than the Hubble radius, leading to oscillatory motion,
whereas this term is negligible on scales larger than the Hubble radius,
and the evolution of $\varphi$ is determined primarily by gravity. Note
that in general cosmological models the Hubble radius is much smaller than the
horizon (the forward light cone calculated from the initial time). In
an inflationary universe, the horizon is larger by a factor of at least 
${\rm exp}(N)$, where $N$ is the number of e-foldings of inflation, and the
lower bound is taken on if the Hubble radius and horizon coincide until
inflation begins. It is very important to realize this difference, a
difference which is obscured in most articles on cosmology in which the
term ``horizon'' is used when ``Hubble radius'' is meant. Note, in
particular, that the homogeneous inflaton field contains causal information
on super-Hubble but sub-horizon scales. Hence, it is completely consistent
with causality \cite{Fabio1}
to have a microphysical process related to the background
scalar matter field lead to exponential amplification of the amplitude
of fluctuations during reheating on such scales, as it does in models
in which entropy perturbations are present and not suppressed during
inflation \cite{BV,Fabio2}. 

There are, however, general relativistic
conservation laws \cite{Traschen} which imply that adiabatic fluctuations
produced locally must be Poisson-statistic suppressed on scales larger than
the Hubble radius. For example, fluctuations produced by the formation of
topological defects at a phase transition in the early universe are
initially isocurvature (entropy) in nature (see e.g. \cite{TTB} for
a discussion). Via the source term in the
equation of motion (\ref{rhbeq4}), a growing adiabatic mode is induced, but
at any fixed time the spectrum of the curvature fluctuation on scales larger
than the Hubble radius has index $n = 4$ (Poisson). A similar conclusion
applies to recently discussed models \cite{Dvali,Kofman} of modulated
reheating as a new source of density perturbations (see \cite{Vernizzi} for
a nice discussion), and to models in which moduli fields obtain masses after
some symmetry breaking, their quantum fluctuations then inducing cosmological
fluctuations. A prototypical example is given by axion fluctuations in an
inflationary universe (see e.g. \cite{ABT} and references therein).

To understand the generation and evolution of fluctuations in current
models of the very early Universe, we need both Quantum Mechanics
and General Relativity, i.e. quantum gravity. At first sight, we
are thus faced with an intractable problem, since the theory of quantum
gravity is not yet established. We are saved by the fact that today
on large cosmological scales the fractional amplitude of the fluctuations
is smaller than 1. Since gravity is a purely attractive force, the
fluctuations had to have been - at least in the context of an eternally
expanding background cosmology - very small in the early Universe. Thus,
a linearized analysis of the fluctuations (about a classical
cosmological background) is self-consistent.

From the classical theory of cosmological perturbations discussed in the
previous section, we know that the analysis of scalar metric inhomogeneities
can be reduced - after extracting gauge artifacts -
to the study of the evolution of a single fluctuating
variable. Thus, we conclude that the quantum theory of cosmological
perturbations must be reducible to the quantum theory of a single
free scalar field which we will denote by $v$. 
Since the background in which this scalar field
evolves is time-dependent, the mass of $v$ will be time-dependent. The
time-dependence of the mass will lead to quantum particle production
over time if we start the evolution in the vacuum state for $v$. As
we will see, this quantum particle production corresponds to the
development and growth of the cosmological fluctuations. Thus,
the quantum theory of cosmological fluctuations provides a consistent
framework to study both the generation and the evolution of metric
perturbations. The following analysis is based on Part II of \cite{MFB}.
 
\subsection{Outline of the Analysis}

In order to obtain the action for linearized cosmological
perturbations, we expand the action to quadratic order in
the fluctuating degrees of freedom. The linear terms cancel
because the background is taken to satisfy the background
equations of motion.

We begin with the Einstein-Hilbert action for gravity and the
action of a scalar matter field (for the more complicated
case of general hydrodynamical fluctuations the reader is
referred to \cite{MFB}) 
\begin{equation} \label{action}
S \, = \,  \int d^4x \sqrt{-g} \bigl[ - {1 \over {16 \pi G}} R
+ {1 \over 2} \partial_{\mu} \varphi \partial^{\mu} \varphi - V(\varphi)
\bigr] \, ,
\end{equation}
where $R$ is the Ricci curvature scalar.

The simplest way to proceed is to work in 
longitudinal gauge, in which the metric and matter take the form
\begin{eqnarray} \label{long}
ds^2 \, &=& \, a^2(\eta)\bigl[(1 + 2 \phi(\eta, {\bf x}))d\eta^2
- (1 - 2 \psi(t, {\bf x})) d{\bf x}^2 \bigr]  \\
\varphi(\eta, {\bf x}) \, 
&=& \, \varphi_0(\eta) + \delta \varphi(\eta, {\bf x}) \nonumber \, .
\end{eqnarray}

The next step is to reduce the number of degrees of freedom. First,
as already mentioned in the previous section, the off-diagonal
spatial Einstein equations force $\psi = \phi$ since
$\delta T^i_j = 0$ for scalar field matter (no anisotropic stresses
to linear order). The two remaining fluctuating variables
$\phi$ and $\varphi$ must be linked by the Einstein constraint
equations since there cannot be matter fluctuations without induced
metric fluctuations. 

The two nontrivial tasks of the lengthy \cite{MFB} computation 
of the quadratic piece of the action is to find
out what combination of $\varphi$ and $\phi$ gives the variable $v$
in terms of which the action has canonical kinetic term, and what the form
of the time-dependent mass is. This calculation involves inserting
the ansatz (\ref{long}) into the action (\ref{action}),
expanding the result to second order in the fluctuating fields, making
use of the background and of the constraint equations, and dropping
total derivative terms from the action. In the context of
scalar field matter, the quantum theory of cosmological
fluctuations was developed by Mukhanov \cite{Mukh2,Mukh3} (see
also \cite{Sasaki}). The result is the following
contribution $S^{(2)}$ to the action quadratic in the
perturbations:
\begin{equation} \label{pertact}
S^{(2)} \, = \, {1 \over 2} \int d^4x \bigl[v'^2 - v_{,i} v_{,i} + 
{{z''} \over z} v^2 \bigr] \, ,
\end{equation}
where the canonical variable $v$ (the ``Mukhanov variable'' introduced
in \cite{Mukh3} - see also \cite{Lukash}) is given by
\begin{equation} \label{Mukhvar}
v \, = \, a \bigl[ \delta \varphi + {{\varphi_0^{'}} \over {\cal H}} \phi
\bigr] \, ,
\end{equation}
with ${\cal H} = a' / a$, and where
\begin{equation} \label{zvar}
z \, = \, {{a \varphi_0^{'}} \over {\cal H}} \, .
\end{equation}
In both the cases of power law inflation and slow roll inflation, 
${\cal H}$ and $\varphi_0^{'}$ are proportional and hence (as long as
the equation of state does not change over time)
\begin{equation} \label{zaprop}
z(\eta) \, \sim \, a(\eta) \, .
\end{equation}
Note that the variable $v$ is related to the curvature
perturbation ${\cal R}$ in comoving coordinates introduced
in \cite{Lyth} and closely related to the variable $\zeta$ used
in \cite{BST,BK}:
\begin{equation} \label{Rvar}
v \, = \, z {\cal R} \, .
\end{equation}

The equation of motion which follows from the action (\ref{pertact}) is
\begin{equation} \label{pertEOM}
v^{''} - \nabla^2 v - {{z^{''}} \over z} v \, = \, 0 \, ,
\end{equation}
or, in momentum space,
\begin{equation} \label{pertEOM2}
v_k^{''} + k^2 v_k - {{z^{''}} \over z} v_k \, = \, 0 \, ,
\end{equation}
where $v_k$ is the k'th Fourier mode of $v$. As a consequence of
(\ref{zaprop}), the mass term in the above equation is given
by the Hubble scale
\begin{equation}
k_H^2 \, \equiv \, {{z^{''}} \over z} \, \simeq \, H^2 \, .
\end{equation}
Thus, it immediately follows from (\ref{pertEOM2}) that on small
length scales, i.e. for
$k > k_H$, the solutions for $v_k$ are constant amplitude oscillations . 
These oscillations freeze out at Hubble radius crossing,
i.e. when $k = k_H$. On longer scales ($k \ll k_H$), the solutions
for $v_k$ increase as $z$:
\begin{equation} \label{squeezing}
v_k \, \sim \, z \,\, , \,\,\,   k \ll k_H \, .
\end{equation}

Given the action (\ref{pertact}), the quantization of the cosmological
perturbations can be performed by canonical quantization (in the same
way that a scalar matter field on a fixed cosmological background
is quantized \cite{BD}). 

The final step in the quantum theory of cosmological perturbations
is to specify an initial state. Since in inflationary cosmology
all pre-existing classical fluctuations are red-shifted by the
accelerated expansion of space, one usually assumes (we will
return to a criticism of this point when discussing the
trans-Planckian problem of inflationary cosmology) that the field
$v$ starts out at the initial time $t_i$ mode by mode in its vacuum
state. Two questions immediately emerge: what is the initial time $t_i$,
and which of the many possible vacuum states should be chosen. It is
usually assumed that since the fluctuations only oscillate on sub-Hubble
scales, the choice of the initial time is not important, as long
as it is earlier than the time when scales of cosmological interest
today cross the Hubble radius during the inflationary phase. The
state is usually taken to be the Bunch-Davies vacuum (see e.g. \cite{BD}),
since this state is empty of particles at $t_i$ in the coordinate frame
determined by the FRW coordinates (see e.g. \cite{RB84} for a
discussion of this point), and since the Bunch-Davies state is
a local attractor in the space of initial states in an expanding
background (see e.g. \cite{BHill}). Thus, we choose the initial
conditions
\begin{eqnarray} \label{incond}
v_k(\eta_i) \, = \, {1 \over {\sqrt{2 \omega_k}}} \\
v_k^{'}(\eta_i) \, = \, {{\sqrt{\omega_k}} \over {\sqrt{2}}} \, \, \nonumber
\end{eqnarray} 
where here $\omega_k = k$, and $\eta_i$ is the conformal time corresponding
to the physical time $t_i$.

Let us briefly summarize the quantum theory of cosmological perturbations.
In the linearized theory, fluctuations are set up at some initial
time $t_i$ mode by mode in their vacuum state. While the wavelength
is smaller than the Hubble radius, the state undergoes quantum
vacuum fluctuations. The accelerated expansion of the
background redshifts the length scale beyond the Hubble radius. The 
fluctuations freeze out when the length scale is equal to the Hubble
radius. On larger scales, the amplitude of $v_k$ increases as the
scale factor. This corresponds to the squeezing of the quantum state present
at Hubble radius crossing (in terms of classical general relativity, it
is self-gravity which leads to this growth of fluctuations). As
discussed e.g. in \cite{PolStar}, the squeezing of the quantum vacuum state
sets up the classical correlations in the wave function of the
fluctuations which are an essential ingredient in the classicalization of the
perturbations.

\subsection{Application to Inflationary Cosmology}

We will now use the quantum theory of cosmological
perturbations developed above to calculate the spectrum
of curvature fluctuations in inflationary cosmology. 

We need to compute the power spectrum ${\cal P}_{\cal R}(k)$ 
of the curvature fluctuation ${\cal R}$ defined in (\ref{Rvar}), namely
\begin{equation}
{\cal R} \, = \, z^{-1} v \, = \, 
\phi + \delta \varphi {{\cal H} \over {\varphi_0^{'}}}
\end{equation}
The idea in calculating the power spectrum at a late time $t$ is
to first relate the power spectrum via the growth rate (\ref{squeezing})
of $v$ on super-Hubble scales to the power spectrum at the time $t_H(k)$
of Hubble radius crossing, and to then use the constancy of the amplitude
of $v$ on sub-Hubble scales to relate it to the initial conditions
(\ref{incond}). Thus
\begin{eqnarray} \label{finalspec1}
{\cal P}_{\cal R}(k, t) \, \equiv  \, k^3 {\cal R}_k^2(t) \, 
&=& \, k^3 z^{-2}(t) |v_k(t)|^2 \\
&=& \, k^3 z^{-2}(t) \bigl( {{z(t)} \over {z(t_H(k))}} \bigr)^2
|v_k(t_H(k))|^2 \nonumber \\
&=& \, k^3 z^{-2}(t_H(k)) |v_k(t_H(k))|^2 \nonumber \\
&\sim& \, k^3 a^{-2}(t_H(k)) |v_k(t_i)|^2 \, , \nonumber
\end{eqnarray}
where in the final step we have used (\ref{zaprop}) and the
constancy of the amplitude of $v$ on sub-Hubble scales. Making use
of the condition 
\begin{equation} \label{Hubble3}
a^{-1}(t_H(k)) k \, = \, H 
\end{equation}
for Hubble radius crossing, and of the
initial conditions (\ref{incond}), we immediately see that
\begin{equation} \label{finalspec2}
{\cal P}_{\cal R}(k, t) \, \sim \, k^3 k^{-2} k^{-1} H^2 \, ,
\end{equation}
and that thus a scale invariant power spectrum with amplitude
proportional to $H^2$ results, in agreement with what was
argued on heuristic grounds in Section (\ref{Sec1}).

\subsection{Quantum Theory of Gravitational Waves}

The quantization of gravitational waves parallels the
quantization of scalar metric fluctuations, but is
more simple because there are no gauge ambiguities. Note
that at the level of linear fluctuations, scalar metric
fluctuations and gravitational waves are independent. Both
can be quantized on the same cosmological background determined
by the background scale factor and the background matter. However, in
contrast to the case of scalar metric fluctuations, the tensor
modes are also present in pure gravity (i.e. in the absence of
matter).

Starting point is the action (\ref{action}). Into this
action we insert the metric which corresponds to a 
classical cosmological background plus tensor metric
fluctuations:
\begin{equation}
ds^2 \, = \, a^2(\eta) \bigl[ d\eta^2 - (\delta_{ij} + h_{ij}) dx^i dx^j 
\bigr]\, ,
\end{equation}
where the second rank tensor $h_{ij}(\eta, {\bf x})$ represents the
gravitational waves, and in turn can be decomposed as
\begin{equation}
h_{ij}(\eta, {\bf x}) \, = \, h_{+}(\eta, {\bf x}) e^+_{ij}
+ h_{x}(\eta, {\bf x}) e^x_{ij}
\end{equation}
into the two polarization states. Here, $e^{+}_{ij}$ and $e^{x}_{ij}$ are
two fixed polarization tensors, and $h_{+}$ and $h_{x}$ are the two 
coefficient functions.

To quadratic order in the fluctuating fields, the action consists of
separate terms involving $h_{+}$ and $h_{x}$. Each term is of the form
\begin{equation} \label{actgrav}
S^{(2)} \, = \, \int d^4x {{a^2} \over 2} \bigl[ h'^2 - (\nabla h)^2 \bigr] 
\, ,
\end{equation}
leading to the equation of motion
\begin{equation}
h_k^{''} + 2 {{a'} \over a} h_k^{'} + k^2 h_k \, = \, 0 \, .
\end{equation}
The variable in terms of which the action (\ref{actgrav}) has canonical
kinetic term is
\begin{equation} \label{murel}
\mu_k \, \equiv \, a h_k \, ,
\end{equation}
and its equation of motion is
\begin{equation}
\mu_k^{''} + \bigl( k^2 - {{a''} \over a} \bigr) \mu_k \, = \, 0 \, .
\end{equation}
This equation is very similar to the corresponding equation (\ref{pertEOM2}) 
for scalar gravitational inhomogeneities, except that in the mass term
the scale factor $a(\eta)$ replaces $z(\eta)$, which leads to a
very different evolution of scalar and tensor modes during the reheating
phase in inflationary cosmology during which the equation of state of the
background matter changes dramatically.
 
Based on the above discussion we have the following theory for the
generation and evolution of gravitational waves in an accelerating
Universe (first developed by Grishchuk \cite{Grishchuk}): 
waves exist as quantum vacuum fluctuations at the initial time
on all scales. They oscillate until the length scale crosses the Hubble
radius. At that point, the oscillations freeze out and the quantum state
of gravitational waves begins to be squeezed in the sense that
\begin{equation}
\mu_k(\eta) \, \sim \, a(\eta) \, ,
\end{equation}
which, from (\ref{murel}) corresponds to constant amplitude of $h_k$.
The squeezing of the vacuum state leads to the emergence of classical
properties of this state, as in the case of scalar metric fluctuations.

\section{Conceptual Problems of Inflationary Cosmology}

After this detailed survey of the theory of cosmological perturbations
applied to inflationary cosmology we can now turn to some conceptual
problems of cosmological inflation, and ways in which string theory
may help address these issues. The first two problems relate to the
cosmological perturbations we have just discussed.

The first problem concerns the amplitude of the cosmological fluctuations.
Considering the simplest large-field potential 
\begin{equation}
V(\varphi) \, = \, {1 \over 2} m^2 \varphi^2 \, ,
\end{equation}
the result (\ref{final}) for the amplitude of the gravitational potential
$\phi$ at late times and large scales (which modulo a factor of order unity
gives the amplitude of the CMB fluctuations on large angular scales and hence
should be of the order $10^{-5}$) yields (making use of the fact that the
result from (\ref{final}) must be evaluated for field values $\varphi \sim 
m_{pl}$ when the relevant scales exit the Hubble radius)
\begin{equation} 
\phi(t_f(k)) \, \sim \, {m \over {m_{pl}}} \, .
\end{equation}
Hence, the value of $m$ must be chosen to be about $10^{13}$GeV, introducing
a new hierarchy problem into particle physics model building. In a model
with quartic potential
\begin{equation}
V(\varphi) \, = \, {1 \over 4} \lambda \varphi^4 \, 
\end{equation}
we obtain a severe constraint on the value of the self coupling constant
($\lambda \ll 10^{-10}$ modulo factors of $2 \pi$), a constraint which implies
that the inflaton cannot be in thermal equilibrium before inflation. To reach
this conclusion we have used ``naturalness'' considerations on coupling constants
which state that the lower bound on the self coupling constant $\lambda$ implies
lower bounds on the coupling constants describing interactions of $\varphi$
with other matter fields, since such interactions generate at higher loop
order contributions to the renormalized value of $\lambda$.
As shown in \cite{Adams}, this hierarchy problem is quite general.

As has recently been shown \cite{Fabio2}, this problem is worse in many
inflationary models with entropy fluctuations. If the entropy perturbations
are not suppressed during inflation, they can be parametrically amplified
during reheating \cite{BKM,Fabio1,BV,Fabio2}. This results in fluctuations
which are nonlinear after inflation independent of the values of
the coupling constants, a result derived including the back-reaction
of these fluctuations in the Hartree approximation \cite{Zibin}. Such models
are thus phenomenologically ruled out.

The second problem is more important and will be discussed at length in the
next section. Basically, since in most scalar field-driven inflationary
models the period of inflation lasts much longer than the minimal number
of e-foldings required in order that scales of current cosmological interest
start out inside the Hubble radius at the beginning of inflation, in such
models these scales thus originate with a wavelength 
much smaller than the Planck length, and hence
the justification for using the formalism of the previous section to
compute the evolution of fluctuations is doubtful. This is the 
{\it trans-Planckian problem} for inflationary cosmology 
\cite{RHBrev} which becomes
the {\it trans-Planckian window of opportunity} for string theory.

Scalar field-driven inflationary models have been shown to be
geodesically incomplete in the past \cite{Borde}. Hence, we know
that this model cannot describe the very early Universe. A major
challenge for string cosmology is to provide this description.

Most importantly, scalar field-driven inflation uses the time-independent
part of $V(\varphi)$ to generate inflation. However, it is 
observationally known that (but theoretically not understood why) 
the time independent quantum vacuum contribution to the energy density
of any quantum field does not gravitate. This is the famous cosmological
constant problem. It may turn out that the solution of the cosmological
constant problem will remove not only quantum vacuum energy but also
the part of $V(\varphi)$ which generates inflation. I view this issue as the
Achilles heel of scalar field-driven inflationary cosmology.

Finally, standard model particle physics does not provide a candidate
for the inflaton. Models of particle physics beyond the standard model
open the window for providing realizations of inflation.

Why can string theory help? First of all, string theory contains
many fields which are massless in the early Universe, namely the moduli
fields. Thus, it provides many candidates for an inflaton. In
addition, it is quite possible that the hierarchy of scales
required to give the right magnitude of density fluctuations emerges
from a hierarchy of symmetry breaking scales in string theory. Secondly,
string theory is supposed to describe the physics on all scales. Thus,
in string cosmology the equations for evolving the fluctuations will
be unambiguous (even if they are not known today), and the trans-Planckian
problem will be solved. One of the main goals of string theory is
to provide a nonsingular cosmology (and a concrete realization of this
goal was provided in \cite{Vafa}). Thus, string theory should be able
to provide a consistent theory of the very early Universe. This theory
might connect with late-time cosmology through a period of inflation,
or through scenarios more similar to \cite{PBB,EKP}.

\section{The Trans-Planckian Window for Superstring Cosmology}

\begin{figure}
\centering
\includegraphics[height=8cm]{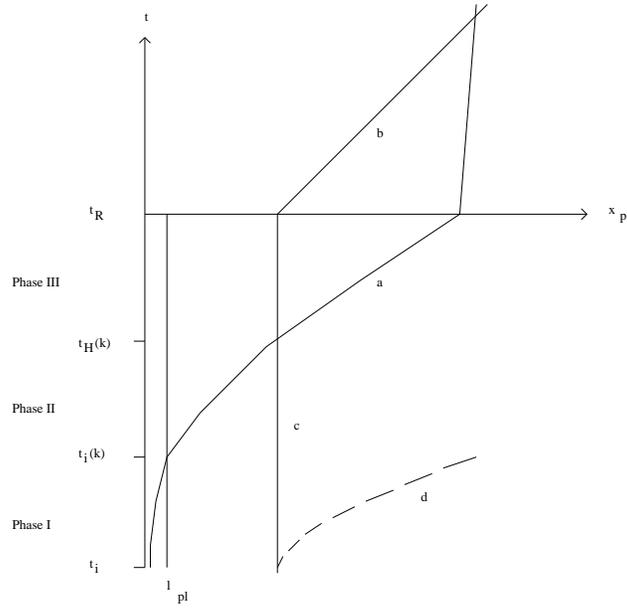}
\caption{Space-time diagram (physical distance vs. time)
showing the origin of
the trans-Planckian problem of inflationary cosmology: at very
early times, the wavelength is smaller than the
Planck scale $\ell _{\rm Pl}$ (Phase I), at intermediate times
it is larger than $\ell _{\rm Pl}$ but smaller than the Hubble
radius $H^{-1}$ (Phase II), and at late times during
inflation it is larger than the Hubble radius (Phase III).
The line labeled a) is the physical wavelength associated
with a fixed comoving scale $k$. The line b) is the Hubble radius
or horizon in SBB cosmology. Curve c) shows the
Hubble radius during inflation. The horizon in inflationary
cosmology is shown in curve d).
\label{FigTP}}
\end{figure}

The same background dynamics which yields the causal generation mechanism
for cosmological fluctuations, the most spectacular success of inflationary
cosmology, bears in it the nucleus of the ``trans-Planckian problem''. This
can be seen from Fig. (\ref{FigTP}). If inflation lasts only slightly
longer than the minimal time it needs to last in order to solve the
horizon problem and to provide a
causal generation mechanism for CMB fluctuations, then the 
corresponding physical wavelength of these fluctuations
is smaller than the Planck length at the beginning of the period of inflation.
The theory of cosmological perturbations is based on classical general
relativity coupled to a weakly coupled scalar field description of
matter. Both the theories of gravity and of matter will break down
on trans-Planckian scales, and this immediately leads to the trans-Planckian
problem: are the predictions of standard inflationary cosmology robust
against effects of trans-Planckian physics \cite{RHBrev}?

This question has recently been addressed using a variety of
techniques. The simplest method is to replace the usual 
dispersion relation for cosmological perturbations by an ad-hoc
modified relation, as was done in \cite{Unruh,CJ} in the context of 
studying the dependence of the thermal spectrum of black hole radiation 
on trans-Planckian physics. We will discuss the application of this
method to cosmology \cite{MB,BM,Niemeyer} below. Other methods include
considerations of modifications of the evolution of cosmological
fluctuations coming from 
string-motivated space-space uncertainty relations 
\cite{Greene,EG1,EG2,Mangano,Hassan}, 
string-motivated space-time uncertainty relations \cite{Ho} (reviewed below), 
from minimal length considerations \cite{KN}, 
from an effective action analysis \cite{Holman}, 
from a minimal trans-Planckian physics viewpoint (starting each mode
out in the vacuum state of the usual action for cosmological perturbations
at the time when the physical wavelength is equal to the new fundamental
length) \cite{Dan,Bozza}, 
and from the point of view of boundary renormalization group analysis
\cite{Schalm}.   

The simplest way of modeling the possible effects of trans-Planckian
physics \cite{MB,BM,Niemeyer} on the evolution of cosmological perturbations, 
while keeping the mathematical analysis simple, is to replace
the linear dispersion relation $\omega _{_{\rm phys}}=k_{\rm phys}$
of the usual equation for cosmological perturbations by
a non standard dispersion relation $\omega _{_{\rm phys}}=\omega
_{_{\rm phys}}(k_{\rm phys})$ which differs from the standard one only
for physical wavenumbers larger than the Planck scale.
This amounts to replacing $k^2$ appearing in
(\ref{pertEOM2}) with $k_{\rm eff}^2(n,\eta )$ defined by
\begin{equation}
k^2 \, \rightarrow \, k_{\rm eff}^2(k,\eta ) \equiv 
a^2(\eta )\omega _{_{\rm phys}}^2\biggl(\frac{k}{a(\eta )}\biggr).
\end{equation}
For a fixed comoving mode, this implies that the dispersion relation
becomes time-dependent. Therefore, the equation of motion of the
quantity $v_k(\eta)$ takes the form (with $z(\eta) \propto a(\eta)$)
\begin{equation} 
\label{eom2}
v_k'' + \biggl[k_{\rm eff}^2(k,\eta ) - {{a''} 
\over a}\biggr]v_k \, = \, 0 \, .
\end{equation}
A more rigorous derivation of this 
equation, based on a variational principle, has been provided \cite{LLMU} 
(see also Ref.~\cite{jacobson}).

The evolution of modes thus must be considered separately in three
phases, see Fig. (\ref{FigTP}). In Phase I the wavelength is smaller
than the Planck scale, and trans-Planckian physics can play
an important role. In Phase II, the wavelength is larger than the
Planck scale but smaller than the Hubble radius. In this phase,
trans-Planckian physics will have a negligible effect
(this statement can be quantified \cite{Shenker}). Hence,
by the analysis of the previous section, the wave function of fluctuations is
oscillating in this phase, 
\begin{equation}
\label{vsubH}
v_k \, = \, B_1\exp(-ik\eta )+B_2\exp(ik\eta )
\end{equation}
with constant coefficients $B_1$ and $B_2$. In the standard approach,
the initial conditions are fixed in this region and the usual choice
of the vacuum state leads to $B_1=1/\sqrt{2k}$, $B_2=0$.  
Phase III starts at the time $t_{\rm H}(k)$ when the
mode crosses the Hubble radius. During this phase, the wave function is
squeezed.

One source of trans-Planckian effects \cite{MB,BM} on observations
is the possible non-adiabatic evolution of the wave function
during Phase I. If this occurs, then it is possible that the
wave function of the fluctuation mode is not in its vacuum state when
it enters Phase II and, as a consequence, the coefficients $B_1$ and
$B_2$ are no longer given by the standard expressions above. In this
case, the wave function will not be in its vacuum state when it crosses
the Hubble radius, and the final spectrum will be different. In
particular, since short wavelength modes spend more time in the 
trans-Planckian phase, it is possible that a deviation of the spectrum
from scale-invariance could be induced in a background which is expanding
almost exponentially. This would be an order one effect on the spectrum
of cosmological perturbations. In
general, $B_1$ and $B_2$ are determined by the matching conditions
between Phase I and II. 
If the dynamics is adiabatic
throughout (in particular if the $a''/a$ term is negligible), the WKB
approximation holds and the solution is always given by
\begin{equation} 
\label{WKBsol}
v_k (\eta )\, \simeq \, \frac{1}{\sqrt{2k_{\rm eff}(k,\eta )}}
\exp\biggl(-i\int _{\eta _{\rm i}}^{\eta }k_{\rm eff}{\rm d}\tau \biggr)
\, ,
\end{equation} 
where $\eta_i$ is some initial time. Therefore, if we start with
a positive frequency solution only and use this solution, we find
that no negative frequency solution appears. Deep in Region II where
$k_{\rm eff} \simeq k$ the solution becomes
\begin{equation}
v_k(\eta ) \simeq {1 \over {\sqrt{2k}}} \exp(-i \phi - i k \eta),
\end{equation}
i.e. the standard vacuum solution times a phase which will disappear
when we calculate the modulus. To obtain 
a modification of the inflationary spectrum, it is necessary to 
find a dispersion relation such that the WKB approximation breaks down 
in Phase I.

A concrete class of dispersion relations
for which the WKB approximation breaks down is \cite{CJ}
\begin{equation}
\label{disprel}
k_{\rm eff}^2(k,\eta ) = k^2 - k^2 \vert b_m \vert
\biggl[{{\ell_{_{\rm pl}}} \over {\lambda(\eta)}} \biggr]^{2m}, 
\end{equation}
where $\lambda (\eta )=2\pi a(\eta )/k$ is the wavelength of a
mode. If we follow the evolution of the modes in Phases I, II and
III, matching the mode functions and their derivatives at the
junction times, the calculation \cite{MB,BM,MB2}
demonstrates that the final spectral index is modified
and that superimposed oscillations appear. It has recently been
shown \cite{Unruh3} that in the case of this class of dispersion relations,
the spectrum of black hole Hawking radiation is also affected.

However, the above example
suffers from several problems. First, in inflationary models with a
long period of inflationary expansion, 
the dispersion relation (\ref{disprel}) leads to complex
frequencies at the beginning of inflation for scales which are of current 
interest in cosmology. Furthermore, the initial conditions for the
Fourier modes of the fluctuation field have to be set in a region
where the evolution is non-adiabatic and the use of the usual vacuum
prescription can be questioned. These problems can be avoided in a toy
model in which we follow the evolution of fluctuations in a bouncing
cosmological background which is asymptotically flat in the past
and in the future. The analysis of \cite{MB3} shows that even in this
case the final spectrum of fluctuations depends on the specific 
dispersion relation used.

\begin{figure}
\centering
\includegraphics[height=6cm]{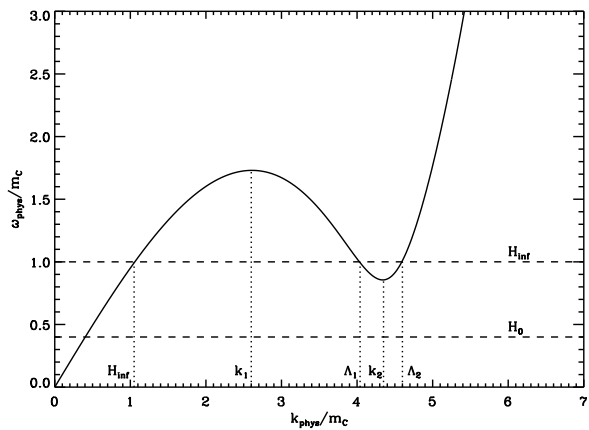}
\caption{Sketch of the dispersion relation of \cite{LLMU}. The adiabaticity
condition on the evolution of fluctuations is broken when the physical
frequency (vertical axis) is smaller than the Hubble expansion rate. During
the phase of inflation, this is the case when the physical wavenumber
(horizontal axis) is in the interval between $\Lambda_1$ and $\Lambda_2$.
Ultraviolet scales are $k > k_0$, and the value $k = k_1$ is the value when
the dispersion relation turns over.
\label{Fig3}}
\end{figure}

An example (see Fig. (\ref{Fig3}))
of a dispersion relation which breaks the WKB approximation in
the trans-Planckian regime but does not lead to the problems mentioned in
the previous paragraph was investigated in \cite{LLMU}. It is a
dispersion relation which is linear for both small and large wavenumbers, but
has an intermediate interval during which the frequency decreases as
the wavenumber increases, much like what happens in (\ref{disprel}).
The violation of the WKB condition occurs for wavenumbers near the local 
minimum of the $\omega(k)$ curve. In this model, modes can be set up
in the far ultraviolet in their Bunch-Davies vacuum. During the time
interval when the physical wavenumber $k_p$ passes through the interval
$\Lambda_1 < k_p < \Lambda_2$ when the physical frequency is smaller than
the value of the Hubble constant $H_{inf}$ in the inflationary phase, the mode
is squeezed and thus is no longer in the local vacuum state when
$k_p < \Lambda_1$.

A justified criticism against the method summarized in the previous analysis
is that the non-standard dispersion relations used are completely ad hoc,
without a clear basis in trans-Planckian physics. 
Other recently explored approaches are motivated by string theory. 
For example, there has been a lot
of recent work 
\cite{Greene,EG1,EG2,Mangano,Hassan} on the implication of space-space
uncertainty relations \cite{Ven,Gross} for the evolution of fluctuations.
The application of the uncertainty relations to the fluctuations lead to
two effects:
Firstly, the equation of motion of the fluctuations
is modified. Secondly, for fixed comoving length scale $k$, the uncertainty
relation is saturated at critical time $t_i(k)$. Thus, in addition
to a modification of the evolution, trans-Planckian physics leads to a
modification of the boundary conditions for the fluctuation modes. The
upshot of this work is that the spectrum of fluctuations is modified. The
magnitude of the deviations can be of the order $H/m_{pl}$, and is thus
in principle measurable.

In \cite{Ho}, the implications of the stringy
space-time uncertainty relation \cite{Yoneya,MY} 
\begin{equation}
\Delta x_{\rm phys} \Delta t \, \geq \, l_s^2 
\end{equation}
on the spectrum of cosmological fluctuations was
studied. Again, application of this
uncertainty relation to the fluctuations leads to two effects. Firstly,
the coupling between the background and the fluctuations is nonlocal in
time, thus leading to a modified dynamical equation of motion
(a similar modification also results \cite{Sera} from quantum
deformations, another example of a consequence of non-commutative basic
physics). Secondly,
the uncertainty relation is saturated at the time $t_i(k)$ when the physical
wavelength equals the string scale $l_s$. Before that time it does not
make sense to talk about fluctuations on that scale. By continuity,
it makes sense to assume that fluctuations on scale $k$ are created at
time $t_i(k)$ in the local vacuum state (the instantaneous WKB vacuum
state).

Let us for the moment neglect the nonlocal coupling
between background and fluctuation, and thus consider the usual
equation of motion for fluctuations in an accelerating background
cosmology.  We assume that $a(t)$ scales as a power of time, i.e. we
consider power-law inflation, and we distinguish two ranges of scales. 
Ultraviolet modes are generated at late times when the Hubble radius is
larger than $l_s$. On these scales, the spectrum of fluctuations does not
differ from what is predicted by the standard theory, since at the time
of Hubble radius crossing the fluctuation modes will be in their vacuum
state. However, the evolution of the infrared modes which are created when
the Hubble radius is smaller than $l_s$ is different. The fluctuations
undergo {\it less} squeezing than they do in the absence of the
uncertainty relation, and hence the final amplitude of fluctuations is
lower. From the equation (\ref{finalspec1}) for the power spectrum of
fluctuations, and making use of the condition
\begin{equation}
a(t_i(k)) \, = \, k l_s 
\end{equation}
for the time $t_i(k)$ when the mode is generated, it follows immediately
that the power spectrum is scale-invariant
\begin{equation} \label{finalb}
{\cal P}_{\cal R}(k) \, \sim \, k^0 \, . 
\end{equation}
In the standard scenario
of power-law inflation the spectrum is red 
(${\cal P}_{\cal R}(k) \sim k^{n-1}$
with $n < 1$). Taking into account the
effects of the nonlocal coupling between background and fluctuation
mode leads \cite{Ho} to a modification of this result: the spectrum
of fluctuations in a power-law inflationary background is in fact blue 
($n > 1$). 

Note that, if we neglect the nonlocal coupling between background and
fluctuation mode, the result of (\ref{finalb}) also holds in a 
cosmological background which is NOT accelerating. Thus, we have a method
of obtaining a scale-invariant spectrum of fluctuations without inflation.
This result was also subsequently obtained in \cite{Wald}, however without a
micro-physical basis for the prescription for the initial conditions.

An approach to the trans-Planckian issue pioneered by 
Danielsson \cite{Dan} which has recently received a lot of
attention is to avoid the issue of the unknown trans-Planckian
physics and to start the evolution of the fluctuation modes at
the mode-dependent time when the wavelength equals the limiting
scale. Obviously, the resulting spectrum will depend sensitively
on which state is taken to be the initial state. The vacuum
state is not unambiguous, and the choice of a state minimizing
the energy density depends on the space-time splitting \cite{Bozza}.
The signatures of this prescription are typically oscillations
superimposed on the usual spectrum. The amplitude of this effect
depends sensitively on the prescription of the initial state,
and for a fixed prescription also on the background cosmology.
For a discussion of these issues and a list of references on
this approach the reader is referred to \cite{Martin03}. Note, in
particular, that for a fixed background cosmology and for a fixed
initial condition prescription, the amplitude of the correction terms
in the spectrum may be different for scalar cosmological perturbations
on one hand and gravitational waves or test scalar matter fields on
the other hand.

If the ultraviolet modes are not in their vacuum state for wavelengths
between the Planck or string scale and the Hubble radius, then the
question of their back-reaction on the background geometry 
\cite{Tanaka,Starob5,Porrati} arises. The naive expectation is that
such ultraviolet modes have the equation of state of radiation, and
that the mode occupation number hence must be very small (of order
$(H_{inf} / m_{pl})^2$) in order for the total energy density carried
by these modes not to prevent inflation. This issue can be analyzed
in the model of \cite{LLMU} in a setting in which there are no
ultraviolet problems (in the naive approach one needs to assume that
modes are continuously generated). In this approach, all modes
start in their vacuum state, they are squeezed when $\Lambda_1 <
k_p < \Lambda_2$, and they oscillate at an excited level thereafter.
In an exponentially expanding background it is clear by time-translation
invariance that the energy density in the ultraviolet modes is constant.
The stress-energy density of a test scalar field with this dispersion
relation has recently been calculated \cite{BM2}, yielding the result
that up to correction terms suppressed by $(H_{inf} / m_{pl})^2$,
the equation of state is that of a cosmological constant. Thus, the
back-reaction of the excited states does not prevent inflation but rather
leads to a renormalization of the cosmological constant. The back-reaction
does effect the speed of rolling of the scalar field, and this yields 
constraints on the mode occupation numbers in the ultraviolet, constraints
which, however, are much weaker than the ones conjectured in
\cite{Tanaka,Starob5,Porrati}.

Another constraint on trans-Planckian physics arises
from the observational limits on the flux of ultra-high-energy
cosmic rays. Such cosmic rays would be produced \cite{Tkachev} in
the present Universe if Trans-Planckian effects would lead to
non-adiabatic mode evolution in the ultraviolet today. In the model
of \cite{LLMU} this does not happen if the present Hubble rate $H_0$
is smaller than the local minimum of the dispersion relation, as in the
situation depicted in Fig. (\ref{Fig3}).

In summary, due to the exponential red-shifting of wavelengths, 
present cosmological
scales originate at wavelengths smaller than the Planck length early on
during the period of inflation. Thus, Planck-scale physics may well encode
information in these modes which can now be observed in the spectrum of
microwave anisotropies. Two examples have been shown to demonstrate the
existence of this ``window of opportunity'' to probe trans-Planckian
physics in cosmological observations. The first method makes use of
modified dispersion relations to probe the robustness of the predictions
of inflationary cosmology, the second applies the stringy space-time
uncertainty relation on the fluctuation modes. Both methods yield the
result that trans-Planckian physics may lead to measurable effects in
cosmological observables.  

\section{Conclusions}

These lectures have focused on the quantum theory of the generation and
evolution of cosmological fluctuations. This is the theory which connects
the fundamental physics of the very early Universe with current cosmological
observations. The general theory was illustrated in the context of
the current paradigm of early universe cosmology, namely the inflationary
universe scenario. In the context of the current approach to
early universe cosmology which involves the coupling of matter described
by quantum field theory to classical general relativity, 
inflationary cosmology 
suffers from important conceptual problems. Resolving these problems
is a major goal for superstring cosmology. Thus, I would argue that
inflationary cosmology requires string theory. 

Whether string theory requires inflation is a different issue. String
theory provides many candidates for an inflaton, and thus it is possible
(maybe even likely) that inflation can be implemented in string theory. 
The study of this issue has recently attracted a lot of attention.
However, one should
keep in mind that current observations do not prove the correctness of
inflation. They show that the spectrum of fluctuations is nearly 
scale-invariant and nearly adiabatic. However, there are other ways
to obtain such a spectrum, although the existing alternatives such
as the ad hoc model mentioned in the previous section of these lecture notes,
or the Pre-Big-Bang \cite{PBB} and Ekpyrotic \cite{EKP} scenarios 
are either not based on fundamental physics or 
not (yet) as developed as inflationary cosmology. Thus, it may be possible
that string theory connects with observations using a cosmological scenario
different than inflation. Even in this case, however, the theory of
cosmological perturbations developed here is applicable. 

A stringy early Universe may involve the dynamics of higher dimensions in
an important way. This is the case for the Ekpyrotic scenario (see
e.g. \cite{TTS,BBP} for some initial work).
In this case, the formalism of cosmological perturbations needs to be
extended (see e.g. \cite{Carsten} for a formalism which is a direct
generalization of what was discussed here). There may be many interesting
effects of bulk fluctuation modes which cannot be seen in a four-dimensional
effective field theory approach. This is a rich field which merits much more
work.

\begin{acknowledgments}
The author wishes to thank the organizers of this Carg\`ese school for the
invitation to lecture and for their hospitality. He also wishes
to thank Pei-Ming Ho and J\'er\^ome Martin for collaborations on the
issue of the trans-Planckian problem of inflationary cosmology.
He is grateful to
the Perimeter Institute for hospitality and financial support during
the time when these lectures were prepared. He is
supported in part by an NSERC Discovery grant (at McGill) and by
the US Department of Energy under Contract DE-FG02-91ER40688, TASK A.

\end{acknowledgments}


\end{document}